\documentclass[useAMS,usenatbib,usegraphicx]{mn2e}
\usepackage{url,rotating,psfrag}

\title[New Insights on the z-$\alpha$ Correlation from Complete Radio Samples]{New Insights on the z-$\alpha$ Correlation from Complete Radio Samples}
\author[L.M. Ker et al.]{L.M. Ker$^{1}$\thanks{E-mail: lmk@roe.ac.uk}, P.N. Best$^{1}$, E.E. Rigby$^{1}$, H.J.A. R\"{o}ttgering$^{2}$, M.A. Gendre$^{3}$.\\
$^{1}$SUPA\thanks{Scottish Universities Physics Alliance}, University of Edinburgh, Institute for Astronomy, Royal Observatory, Edinburgh, EH9 3HJ, UK\\
$^{2}$Leiden Observatory, P.O. Box 9513, 2300 RA, Leiden, The Netherlands\\
$^{3}$Jodrell Bank Centre for Astrophysics, University of Manchester, Alan Turning Building, Oxford Road, Manchester, M13 9PL, UK.}
\begin{document}

\date{Accepted 2011 November 18. Received 2011 November 18; in original form 2011 October 14}

\pagerange{\pageref{firstpage}--\pageref{lastpage}} \pubyear{2011}

\maketitle

\label{firstpage}

\begin{abstract}
The existence of a correlation between observed radio spectral index and redshift has long been used as a method for selecting high redshift radio galaxy candidates. We use 9 highly spectroscopically complete radio samples, selected at different frequencies and flux limits, to determine the efficiency of this method, and compare consistently observed correlations between spectral index ($\alpha$), luminosity (P), linear size (D) and redshift (z) in our samples. We observe a weak correlation between z and $\alpha$ which remains even when Malmquist bias is removed. The strength of the z-$\alpha$ correlation is dependent on both the k-correction and sample selection frequency, in addition to the frequency at which $\alpha$ is measured, and consistent results for both high and low frequency selected samples are only seen if analysis is restricted to just extended radio galaxies. This fits with the popular interpretation that the spectra steepen with z because the radio lobes work against a denser IGM environment as z increases, out to z$\sim$2-3. However we also note that the majority of sources known at z$>$4 are very compact and often display a negatively curved or peaked spectrum, indicative of youth or merger activity, and therefore the low frequency radio spectrum as a whole should be determined; this is something for which the new LOw Frequency ARray will be crucial. We quantify both the efficiency and the completeness of various techniques used to select high-z radio candidates. A steep-spectrum cut applied to low-frequency selected samples can more than double the fraction of high-z sources, but at a cost of excluding over half of the high-z sources present in the original sample. An angular size cut is an almost as equally effective radio-based method as a steep-spectrum cut for maximising the high-z content of large radio samples, and works for both high and low frequency selected samples. In multi-wavelength data, selection first of infrared-faint radio sources remains by far the most efficient method of selecting high-z sources. We present a simple method for selecting high-z radio sources, based purely on combining their observed radio properties of $\alpha$ and angular size,  with the addition of the \emph{K}-band magnitude if available.
\end{abstract}


\section{Introduction}
Vast amounts of multi-frequency radio data at long wavelengths will soon begin to flow from next generation radio instruments such as the LOw Frequency ARray (LOFAR) and eventually the Square Kilometre Array (SKA). With this, opportunities will arise for studying some of the earliest radio sources in the universe, their environments and their evolution over cosmic time. There is also the tantalizing possibility of studying conditions within the Epoch of Reionisation itself through high-z radio sources: if sufficiently bright radio sources can be found at redshifts greater than 6.5, it should be possible to measure absorption signatures of neutral hydrogen, and hence trace changes in the ionisation state of the Universe with cosmic time (e.g. \citealt{CarilliAstron.J.133:2841-28452007}; \citealt{Meiksin2011}). 

The existence of a correlation between redshift and observed spectral index $\alpha$\footnote{where S$_{\nu}\propto \nu^{\alpha}$} for powerful radio sources was first suggested by \citet{Tielens1979} who found that the introduction of a steep spectral index cut led to an increasing fraction of sources without optical counterparts identified on POSS-\emph{I} (\emph{R} $<$ 20) plates. \citet{BlundellAstron.J.117:677-7061999} used the combination of complete samples from the 3CRR \citep{Laing1983}, 6CE \citep{Rawlings2001} and 7CRS (\citealt{Willott2003}; \citealt{Lacy1999}) surveys to confirm that the high frequency (5 GHz) rest-frame spectral index correlates with redshift, but showed that this correlation is weaker for spectral index measured at lower rest-frame frequencies. Since then, there have been many surveys designed to pick out only Ultra Steep Spectrum (USS) sources for further infra-red imaging and spectroscopic follow-up (e.g. \citealt{Rottgering1996}; \citealt{Breuck2000}), having varying degrees of success selecting high-z sources. Many of these have additional selection criteria such as small angular size and faint infrared magnitude applied after the USS cut, which makes it difficult to determine the extent to which the USS cut is responsible for selecting high-z sources.

\citet{KlamerMon.Not.Roy.Astron.Soc.371:852-8662006} present a sample of USS selected galaxies selected from the Sydney University Molonglo Sky Survey (SUMSS), and discuss the apparent mechanisms for the z-$\alpha$ correlation in detail. They dismiss the possibility of k-corrections being the cause of the observed steepening radio spectra, given that the majority of their radio spectra show no evidence for curvature. They suggest that enhanced spectral aging due to inverse Compton losses against the Cosmic Microwave Background (CMB) at high redshift is the most likely origin for the observed z-$\alpha$ correlation,  or that it may arise due to an intrinsic relation between low frequency $\alpha$ and radio luminosity coupled to a Malmquist bias. Following on from this work, \citet{Bryant2009} compare the median redshift of several complete samples with the median redshift obtained from USS selected samples. They find it to be lower in complete samples, and argue that this is strong evidence for the efficiency of the USS technique. However whilst USS samples clearly do select higher redshift sources, sample comparisons between USS and complete samples are not ideal for either optimising or quantifying the efficiency of the technique. The most rigorous approach would be to apply selection criterion to complete samples with significant numbers of sources at the highest redshifts currently known, and quantify the number of high redshift sources included/missed.
\begin{figure*} 
\begin{center}
                \includegraphics[width=\columnwidth]{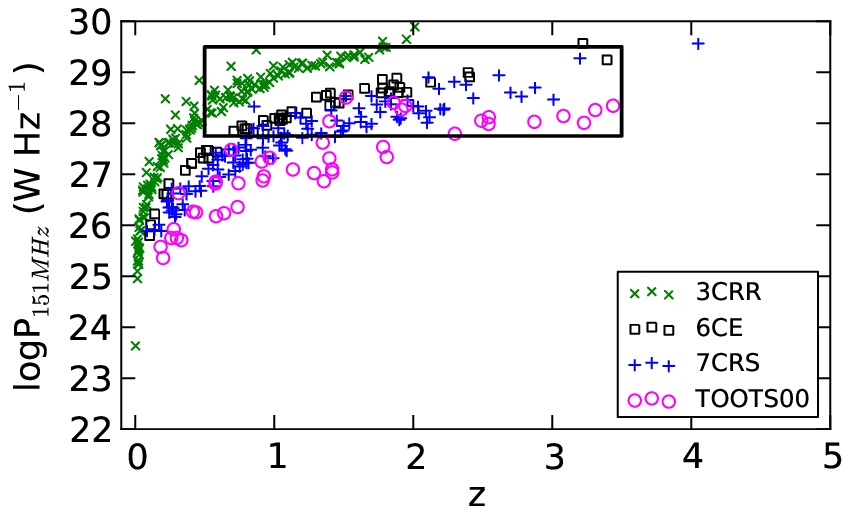}
		\includegraphics[width=\columnwidth]{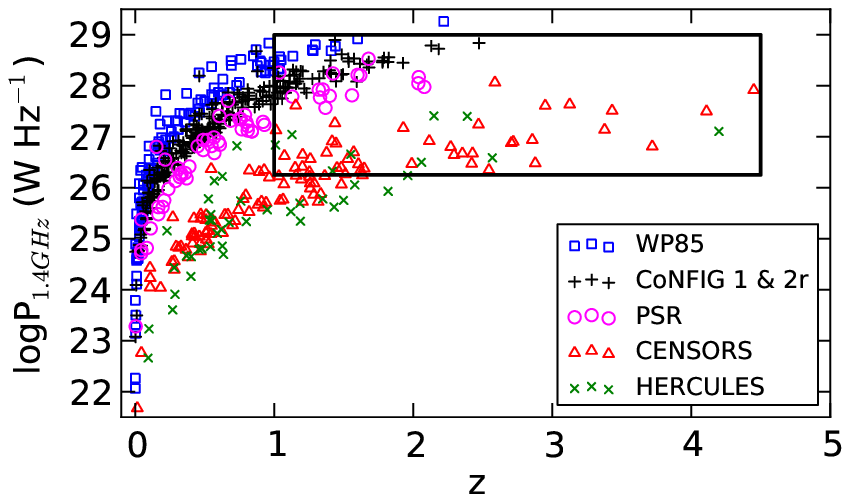}
\caption{\label{Fig1}\small Radio luminosity vs redshift plane coverage for the low frequency selected samples 3CRR, 6CE, 7CRS and TOOTS (left) and for the high frequency selected samples WP85r, PSRr, CoNFIG 1 \& 2r, CENSORS and Hercules (right). Only sources with $\alpha$ $<$ $-$0.5, as used in this study are shown. The boxes indicate the Malmquist-bias-free sections of the P-z plane at high and low frequency, which we use in later analysis, for 151 MHz samples: logP=27.75-29.5, z=0.5-3.5, and for 1.4 GHz samples: logP=26.25-29.0, z=1.0-4.5.}
\end{center}
\end{figure*}

\setcounter{table}{0}
\begin{table*}
\caption{\label{Tab1} Details of the complete samples used in this study. Note that the spectral indices for CoNFIG are taken from Gendre et al. (2010), from a linear fit to flux densities between 1.4 GHz and 178 MHz, rather than a two point spectral index. }
\begin{tabular}{|l|l|l|l|l|l|l|l|}
\hline
  \multicolumn{1}{|c|}{Survey} &
  \multicolumn{1}{c|}{Selection $\nu$ (MHz)} &
  \multicolumn{1}{c|}{no. of Sources} &
  \multicolumn{1}{c|}{Sky Area (sr)} &
  \multicolumn{1}{c|}{Flux Limit (Jy)} &
  \multicolumn{1}{c|}{\% z$_{spect}$} &
  \multicolumn{1}{c|}{\% no z} &
  \multicolumn{1}{c|}{$\alpha$ range (MHz)} \\	
\hline
\hline
3CRR & 178 & 173 & 4.239 & S$_{178}>$10.9 & 100 & 0.0&750-151\\
6CE & 151 & 58 & 0.102 & 2.00$<$S$_{151}<$3.93 & 97 & 3.0 &1400-151 \\
7CI & 151 & 37 & 0.0061 & S$_{151}>$0.51 &90 & 0.0 & 1400-151\\
7CII & 151 & 37 & 0.0069 & S$_{151}>$0.48 &90 & 0.0 &1400-151\\
7CIII & 151 & 54 & 0.009 & S$_{151}>$0.50 &95 & 0.0 & 1400-151\\
TOOTS-00 & 151 & 47 & 0.0015 & S$_{151}>$0.10 & 85 &2.0 & 1400-151\\
WP85r & 1400 & 138 & 9.81 & S$_{1400}>$4 & 95 & 0.0 &5000-1400\\ 
CoNFIG1 & 1400 & 273 & 1.5 & S$_{1400}>$1.3 & 83 & 3.6 & 1400-178\\
CoNFIG2r & 1400 & 61 & 0.89 & 1.0$<$S$_{1400}<$1.3 & 52 & 4.9 & 1400-178\\
PSRr & 1400 & 59 & 0.075 & S$_{1400}>$0.36 & 61 & 0.0 & 2700-1400\\ 
CENSORS & 1400 & 135 & 0.0018 & S$_{1400}>$0.0072 & 78 & 3.7 & 1400-325\\
Hercules & 1400 & 64 & 0.00038 & S$_{1400}>$0.002 & 66 & 3.0 & 1400-610\\
\hline
\end{tabular}
\end{table*}
Despite the wide use of the z-$\alpha$ technique to select high redshift galaxies, there has been very little work on quantifying the efficiency. \citet{Pedani2003} states that, for the first time, they present the true quantitative searching efficiency for high-z radio galaxies using a sample selected from the Molonglo Reference Catalogue (MRC). They utilise 225 sources with full redshift information from this sample to measure the efficiency of optical, USS and size selection. They find that the efficiency (defined as the fraction of z$>$2 sources in the recovered sample) of the USS criterion alone is 0.33, increasing to a maximum of $\sim$0.59 in combination with an optical cut. However, their 225 source sample is not complete, being composed of only objects with redshift information amongst the complete MRC 1 Jy radio sample of 446 sources. They argue that the redshift-complete subsample is representative of the full sample, as both contain similar proportions of USS sources. However they also note that the median magnitude of the subset of galaxies without redshifts is fainter than that for those included. This means firstly, that there is an optical magnitude bias towards brighter magnitudes in the analysed sample, and secondly, that the work is based on the implicit assumption that the USS criterion is more important than optical magnitude in selecting high redshift candidates. With 50\% of the sample not analysed, and at fainter optical magnitudes, the redshift incompleteness towards higher redshift cannot be quantified, and this could be substantial. 

Potential biases such as these are common in the literature, due to the difficulty and expense of building spectroscopically complete radio samples. As such, any attempt to use large collections of radio data available in the literature to investigate evolution of basic radio properties is not valid, despite the large number statistics, without clearly defined and well understood selection criteria. For example, recent work by \citet{KhabibullinaAstrophys.Bull.64No32632009} uses a large sample of 2442 radio galaxies with measured redshifts selected from large publically available radio source catalogues. They determine the dependence of $\alpha$ on z, and select a sample of distant objects using this relation. Crucially however, as they note, the samples they use are not complete in any sense, and some of the largest high-z radio source samples with radio spectra publically available are ones with an USS criterion applied (e.g. \citealt{Breuck2000}) which will irrevocably bias spectral index studies of any sample contructed from them.

In summary, although the existence of the so-called z-$\alpha$ correlation has been known for some time, there has been little attempt to quantify the strength of this consistently across a wide range of spectroscopically complete samples at different selection frequencies, and measure the resultant efficiency of using an USS $\alpha$ cut-off in order to isolate high-z candidates. In this study, we address these shortcomings, thus providing a vital tool for the design of further high-z source searches from upcoming radio surveys by new survey instruments, e.g. LOFAR. This work builds significantly on current knowledge in five ways:

\begin{itemize}
 \item We use nine highly spectroscopically complete and unbiased radio source samples. Most have a spectroscopic completeness in the range 80-100\%, and robust redshift estimates (e.g. photometric or based on the \emph{K}-z relation) are available for the vast majority of the remaining sources, such that all samples are at least 95\% redshift-complete.
 \item We use new radio data from the CENSORS radio sample (\citealt{Best2003}), which contains a large number of sources with z $>$ 2, improving high redshift statistics.
 \item Selection frequency effects are fully explored - four samples are selected at frequencies below 200 MHz, and five at 1.4 GHz.
 \item The samples have a wide range of flux density limits, so that correlations such as the P-$\alpha$ and z-$\alpha$ relations may be safely disentangled.
\item We also consider radio linear size (D) in order to investigate its role in selecting high redshift sources.
\end{itemize}

The layout of this paper is as follows. In Section 2 we describe the complete radio samples used in this study. In Section 3 we give a brief summary of radio source properties and sample selection effects. In Section 4 we investigate observable trends and employ principle component analysis to identify fundamental correlations in the PD$\alpha$z parameter space for various collections of samples. In Section 5 we attempt to fit various functional forms for $\alpha$ to the observed data, and identify large intrinsic scatter in $\alpha$, not dependent on P, z or D. In Section 6 we discuss the physical origins of the observed z-$\alpha$ correlation, and finally in Section 7 we discuss the implications of our findings in the search for the highest redshift radio galaxies, and use complete samples to explore the efficiency of often used techniques in the literature to find these. 

Throughout this work a $\lambda$CDM cosmology is assumed, $\Omega_{\lambda}=0.7$, $\Omega_{M}=0.3$ and $H_{0}$=70 km s$^{-1}$ Mpc$^{-1}$, and magnitudes are in the Vega system.
\vspace{-0.25cm}
\section{Complete Radio Samples Selection}
We want to quantify the z-$\alpha$ correlation at a wide range of frequencies and flux density limits, determine to what extent this is an intrinsic property of sources (rather than being driven by, for example, a P-$\alpha$ correlation) and understand any selection effects present. In order to do this, we collate data from several complete samples already available in the literature: the 3CRR, 6CE, 7CRS and TOOTS-00 selected at low frequency, and the WP85r, CoNFIG1\&2, PSRr, CENSORS, and Hercules samples selected at high frequencies. These samples are described below, summarised in Table \ref{Tab1}, and displayed on the P-z plane in Figure \ref{Fig1}. 
\subsection{The 3CRR Sample}
The 3CRR, or Third Cambridge Revised Revised sample of extragalactic radio sources \citep{Laing1983}, is a complete sample containing all radio sources above 10.9 Jy at 178 MHz in an area of sky covering 4.239 sr. The sample comprises 173 objects in total, and is 100\% spectroscopically complete. The data were obtained from the 3CRR catalogue webpage maintained online\footnote{\url{http://3crr.extragalactic.info/}}. As this sample is the only low frequency selected sample observed at 178 MHz as opposed to 151 MHz, the flux densities are converted to 151 MHz fluxes assuming the spectral indices given in the catalogue. The observed spectral index $\alpha$ is measured between 750 MHz and 178 MHz for this sample.
\subsection{The 6CE Sample}
The 6CE sample is based on a original sample selected by \citet{Eales1985} from the Sixth Cambridge radio survey (6C) comprising of 67 radio sources between 2.2 Jy and 4 Jy selected at 151 MHz, over a sky area of 0.102 sr. For this study, we use a reselected, updated version,  available online\footnote{\url{http://www-astro.physics.ox.ac.uk/~sr/6ce.html}}, the 6CE sample of \citet{Rawlings2001}. This consists of all sources with a 151 MHz flux density in the range 2.00 $\geq$ S$_{151}$ $\geq$ 3.93 Jy in the same 0.102 sr patch of sky. There are 59 sources in total, with all but one having a firm identification, and 56 of the 59 having spectroscopic redshifts. Of the three sources without spectroscopic redshifts, one is obscured by a bright star and so is excluded from the sample, and the other two have a redshift estimate from the \emph{K}-z relation. Observed spectral indices have been calculated between 1.4 GHz and 151 MHz using 1.4 GHz fluxes obtained from the NVSS.

\subsection{The 7CRS Sample}
The 7CRS, or Seventh Cambridge Redshift Survey is composed of three subsamples, 7CI, 7CII and 7CIII. 7CI and 7CII are each composed of 37 sources with flux density limits of S$_{151}\geq$0.51 Jy and S$_{151}\geq$0.48 Jy respectively in the 7C survey and are defined in \citet{Willott2003}. The redshifts and linear sizes for the 7CI \& 7CII samples are available online from the data of \citet{Grimes2004}\footnote{\url{http://www-astro.physics.ox.ac.uk/~sr/grimes.html}}. The 7C-III sample contains 54 objects with a flux limit of S$_{151}>$0.50 Jy, detailed in \citet{Lacy1999}. We utilise the 7CIII data from Table 8 in \citet{Lacy1999} to get the luminosities, redshifts and linear sizes of the sample. 

The spectral indices for this sample are not yet available in a collective form in the literature. We estimate the observed spectral index by cross-matching 151 MHz fluxes for each source from the 7C 151 MHz catalogue of \citet{2007MNRAS.382.1639H} with the NRAO VLA Sky Survey (NVSS; \citealt{Condon1998}), at 1.4 GHz. We checked all extended sources listed in the 7C Hales catalogue as having separate components, and cross checked maps at 151 MHz with NVSS maps in order to correctly identify components and catalogue the correct integrated flux for each source. We have also matched the source list with the  TEXAS 365 MHz/WENSS 327 MHz surveys, the 5C 408 MHz survey, and finally the VLA Low Frequency Sky Survey at 74 MHz (VLSS; \citealt{Cohen2007}) with the addition of the 38 MHz 8C survey for 7CIII \citep{Lacy1999}, all of which are of comparable resolution, for later curvature analysis (see Section 6, and Ker et al, in preparation). We note that it is possible that very extended sources may not have correct fluxes in these catalogues. 

7CI and 7CII both have 90\% spectroscopic redshift completeness, and 7CIII is 95\% complete. The remaining sources in all three subsamples have photometric redshifts estimated from the \emph{K}-z relation.

\subsection{The TOOTS-00 Sample}
The TOOT00 region, \citep{Vardoulaki2009}, is the first complete region of the Tex-Ox-1000 redshift survey of radio sources. This survey selects all sources above 100 mJy in the Cambridge 7C 151 MHz survey, and is designed to be approximately 5 times fainter than the 7CRS, with much greater numbers. \citet{Vardoulaki2009} present complete radio, near-infrared and spectroscopic data or redshift estimates for the first region of the survey, comprising 47 sources. 40 of the radio sources have spectroscopic redshifts, with a further six using a redshift estimated from the \emph{K}-z relation. The final source has a \emph{K}-limit only and we adopt the lower redshift limit as the redshift estimate for this source (the \emph{K}-band data reaches sufficient depth to place the source at high-z, and hence for the lower redshift limit to be adopted as the redshift estimate with little loss of accuracy). The observed spectral index was calculated for each source using flux data at 151 MHz and 1.4 GHz (NVSS) from \citet{Vardoulaki2009}.

\subsection{The Wall and Peacock 2.7 GHz Sample}
The original \citet{1985MNRAS.216..173W} 2.7 GHz 233 source sample covers 9.81 sr of sky, and includes all radio sources brighter than 2 Jy. The sample now stands at  98\% spectroscopically complete (\citealt{Rigby2011}). In this study, we use the 138 source sample, reselected by \citet{Rigby2011} to be complete at 1.4 GHz to a flux limit of 4 Jy with a spectral index between 5 and 1.4 GHz steeper than $-$0.5. This re-selected sample is 97\% spectroscopically complete, and the remaining three sources have photometric redshift estimates. We refer to this sample as WP85r.

\subsection{The CoNFIG Samples}

We utilise two complete samples from the Combined NVSS-FIRST Galaxy catalogue (CoNFIG), ConFIG regions 1 and 2 \citep{Gendre2010}. 

CoNFIG1 contains 273 sources complete to 1.4 GHz $\geq$ 1.3 Jy, and is 83\% spectroscopically complete. In CoNFIG 1, 226 sources have spectroscopic redshifts, 37 have photometric redshift estimates and 10 sources (4\%) have only lower redshift limits from SDSS \emph{I}-band non-detections. These non-detections are not sufficently deep to provide a useful constraint on the redshift (the SDSS limiting \emph{I}-band magnitude only constrains each source to z$\geq$1). However of these 10 sources, only four have an observed spectral index steeper than $-$0.5, and hence should be included in the analysis (see Section 3). We choose not to include these four sources as the redshift estimate is not reliably constrained, and such a small fraction will have a statistically insignificant effect on the results. All four sources have very different morphological types and spectral indices, so are unlikely to be biased toward any one redshift range.

CoNFIG2 contains 132 sources and is complete between 1.3 Jy and 0.8 Jy at 1.4 GHz (only sources with 1.4 GHz fluxes less than 1.3 Jy were used from CoNFIG2, to ensure no duplication with sources also in CoNFIG1). At fainter flux densities the redshift completion of CoNFIG2 is relatively low, so we reselect the sample to above 1 Jy at 1.4 GHz, creating a new sample of 61 sources which we refer to as CoNFIG2r. For this revised sample, reselecting to above 1 Jy at 1.4 GHz reduces the proportion of  unidentified redshift sources to a negligible number of 3. Of these three sources, only two have a spectral index steeper than $-$0.5, and hence should be included in the analysis. As these two have greatly differing spectral indices and morphologies, again they are unlikely to be limited to any one redshift range, and similarly to CoNFIG1, we do not include these two sources in the subsequent analysis. 

The observed spectral index is taken from \citet{Gendre2010}, calculated using a linear fit to flux data points between 1.4 GHz and 151 MHz. We cross-matched the CoNFIG catalogues with the VLSS and 7C 151 MHz \citep{2007MNRAS.382.1639H} catalogue, giving flux data at 74, 151, 365, 408, 1400, 2700 and 5000 MHz for both samples to allow curvature analysis (see Section 6, and Ker et al, in prep).  Frequency coverage for this dataset is very good: only 28 sources have no data at 74/151 MHz, and only 15 have no 2.7/5 GHz data. 

\subsection{Parkes Selected Regions}
The original Parkes Selected Regions (\citealt{Wall1968}; \citealt{1986MNRAS.218...31D}; \citealt{Dunlop1989}) is a complete 178 object sample containing all radio sources brighter than 0.1 Jy over a 0.075 sr sky area at 2.7 GHz. We have reselected the sample at 1.4 GHz $>$ 0.36 Jy as PSRr, to which flux limit there are 59 sources with an observed spectral index between 2.7 GHz and 1.4 GHz steeper than $-$0.5. 36 of these sources have spectroscopic redshifts, with the remaining 23 having redshift estimates from the \emph{K}-z relation or photometric spectral fitting.

\subsection{The CENSORS Sample}
The Combined EIS-NVSS Survey of Radio Sources, or CENSORS sample is a 135 source sample of all radio sources with an NVSS 1.4 GHz flux density greater than 7.2 mJy in a six square degree patch of the sky centred on the ESO Imaging Survey (EIS) Patch D \citep{Best2003}. The sample currently stands at 96\% identified and  78\% spectroscopically complete, and is currently one of the largest highly spectroscopically complete faint 1.4 GHz selected samples in existence (\citealt{Brookes2006}; \citealt{Brookes2008}; \citealt{Rigby2011}; Ker et al. in prep). 105 sources have spectroscopic redshifts, and of the remaining 30, 25 have redshift estimates based on the \emph{K}-z relation, and 5 have only a lower limit redshift estimate from a \emph{K}-band non-detection. At \emph{K}-band limits of $\sim$19 and above, the non-detections are sufficiently deep that the sources must be at high redshift, and the lower redshift limit can be adopted as the estimated redshift without great loss of accuracy. The observed spectral index is measured between 1.4 GHz and 325 MHz (see Ker et al, in prep for 325 MHz data).

\subsection{The Hercules Sample}
The Hercules sample is taken from a field in the  Leiden-Berkeley Deep Survey, and consists of 64 sources selected to have a flux density greater than 2 mJy at 1.4 GHz \citep{Waddington2001}. The spectroscopic completeness stands at 66\%, with 20 sources having photometric redshifts based on the \emph{K}-z relation, and the final two having a redshift limit estimated from \emph{K}-band limits. Again, at \emph{K}-band limits of 20.7 and 19.85 mag respectively, the non-detections are sufficiently deep that the sources must be at high redshift, and the lower redshift limit can be adopted as the estimated redshift. Observed spectral indices are calculated between 1.4 GHz and 610 MHz.

\section{Radio Source Properties and Sample Selections}
\begin{figure*}
\begin{center}
                \includegraphics[width=\textwidth]{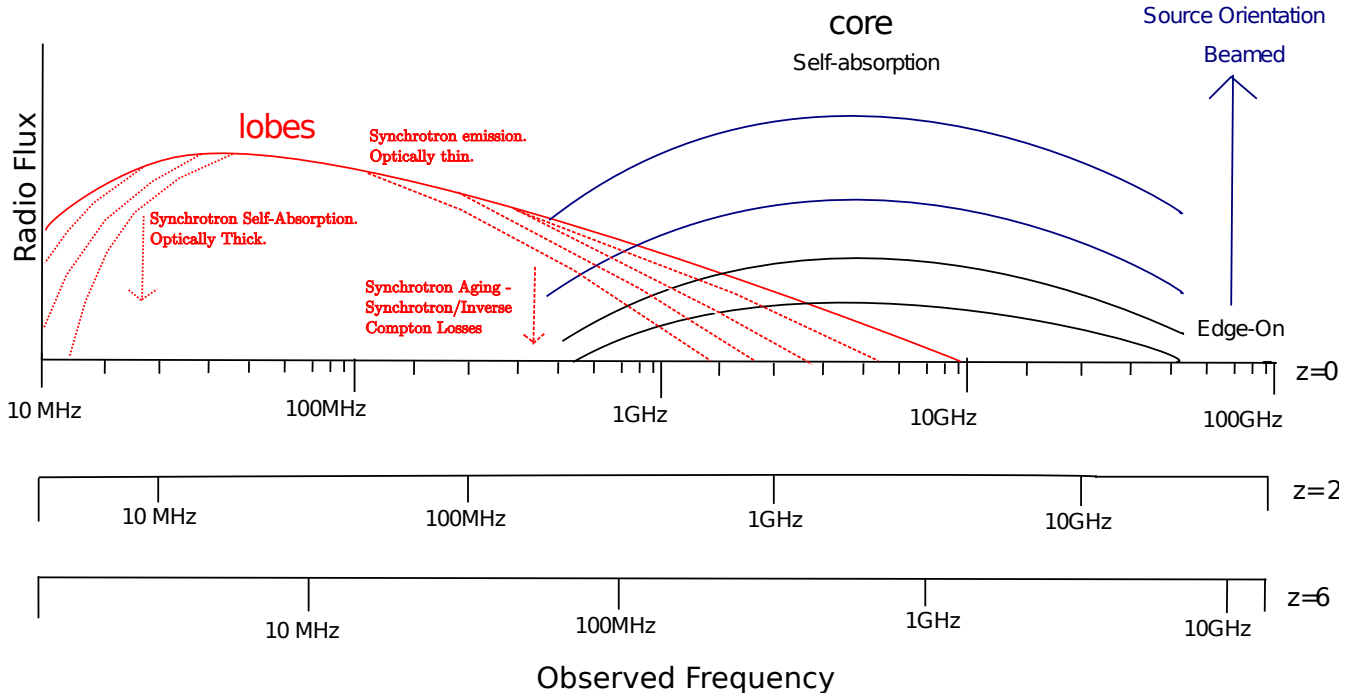}
\caption{\label{Fig2} \small A sketch of the contributions of the various components of a typical double radio galaxy to the rest-frame radio regime, and the effects of synchrotron self-absorption, losses and beaming on the observed radio spectrum. Also shown are the observed frequencies at which these features would be observed at redshifts two and six for comparison.}
\end{center}
\end{figure*}
Complete radio samples will select very different populations, depending on the frequency at which they are selected, and their flux density limit, as different physical contributions dominate at differing rest-frame frequencies. In this study, we utilise samples selected in both the MHz and GHz regimes, representative of existing spectral index studies.

Figure \ref{Fig2} illustrates the contributing components of a typical extended radio galaxy. If the source is not highly beamed, i.e. not viewed along the jet axis, the emission is dominated at low frequency by synchrotron emission in the radio lobes. Radio lobes typically display a steep-spectrum power law slope, which can steepen further at higher frequency due to both synchrotron and inverse Compton losses (see for example \citet{1991ApJ...383..554C} who analyse the radio spectrum of the well-studied local radio galaxy Cygnus A in depth). At low frequencies, the lobe spectrum can turn over due to synchrotron self-absorption. The frequency at which this happens depends on both the size and intensity of the emitting component: it occurs at higher frequencies for smaller emitters, leading to the smallest radio sources at sub-kpc size being GHz Peaked Sources (GPS).

At higher frequencies (above a few GHz), the contribution to the spectrum from the core is often important. Emission from the core is typically flat spectrum, due to the superposition of self-absorbed components of different sizes at the base of the radio jet. If the jet is orientated towards us, it can be Doppler-boosted by beaming, and can become dominant at lower frequencies.

As can be seen from Figure \ref{Fig2}, if a sample is selected at a few hundred MHz, up to high redshifts the radio emission will still be probing the lobe-dominated regime, giving a sample of similar, directly comparable sources. However, if a sample is selected at GHz frequencies or above, sources with a significant core component that are orientated such that the jet is aligned along the line of sight towards us (beamed) will be preferentially included, especially at higher redshifts. 

It is thought that the observed z-$\alpha$ relation may arise because sources at higher redshift have lobes doing work against a denser medium. Working against a denser medium means there will be less adiabatic expansion losses, and therefore greater synchrotron losses, with the result that the source is brighter but the radio spectrum steepens faster. However, as shown, in GHz selected samples, the observed spectral index may alternatively be flattened at the highest redshifts by an increasing contribution of a core component, and be less affected by environment.

Although only rest-frame spectral indices should have any direct physical correlation with other observables (observed spectral indices being a good approximation), as far as possible, we utilise a traditional two point observed spectral index, so as to match the situation most widely encountered in the literature and most simply provided by the observations, e.g. for the selection of USS sources. 

There is a strong argument to exclude as far as possible all significantly beamed sources identified in the samples, as not only will their spectral index estimates be distorted (the beamed component generally being flatter spectrum), they will also be heavily foreshortened in size. However, as our primary motivation is measuring the efficiency of radio-based correlations in selecting high redshift radio sources from radio surveys, we must adopt a simple approach to removing these that can be widely applied to blind radio surveys. In most comparible observational studies, a cut of $\alpha$=$-$0.5 is used as a division, to separate out flat and steep spectrum sources (and indeed such a cut has already been applied in the definition of some of the samples we use). Hence, in order to analyse comparable parts of the radio spectrum, we restrict analysis in this study only sources with an observed spectral index less than $-$0.5; these will largely be of a similar type (lobe dominated). Sources with a flatter spectral index represent a composite population: as well as quasars and core-dominated sources, they may also include for example young peaked radio sources (see analysis in Ker et al, in prep). We do not remove starburst galaxies, as their numbers are negligibly low in all samples. 

Luminosities were calculated for each sample using P$_{\nu}$ = 4$\pi$S$_{\nu}$(1+z)$^{-1-\alpha}$D$_{L}^{2}$, where $\alpha$ is the observed spectral index, defined as $S_{\nu} \propto \nu^{\alpha}$ and D$_{L}$ is the luminosity distance. The transverse linear size in Mpc of each radio source was calculated using $D = \theta D_{A}$ where $\theta$ is the maximum measured angular extent of the radio source on the sky in radians, and $D_{A}$ is the angular diameter distance ($D_{A} = D_{L}/(1+z)^2$). For Hercules and CENSORS $\theta$ is determined at 1.4 GHz; for TOOTS, 7CRS, 6CE and 3CRR $\theta$ is measured at 151 MHz. There are no readily available angular size measurements in the literature for WP85r, PSRr and CoNFIG 1 \& 2r.
\begin{figure*}
\begin{center}
                \includegraphics[width=\textwidth]{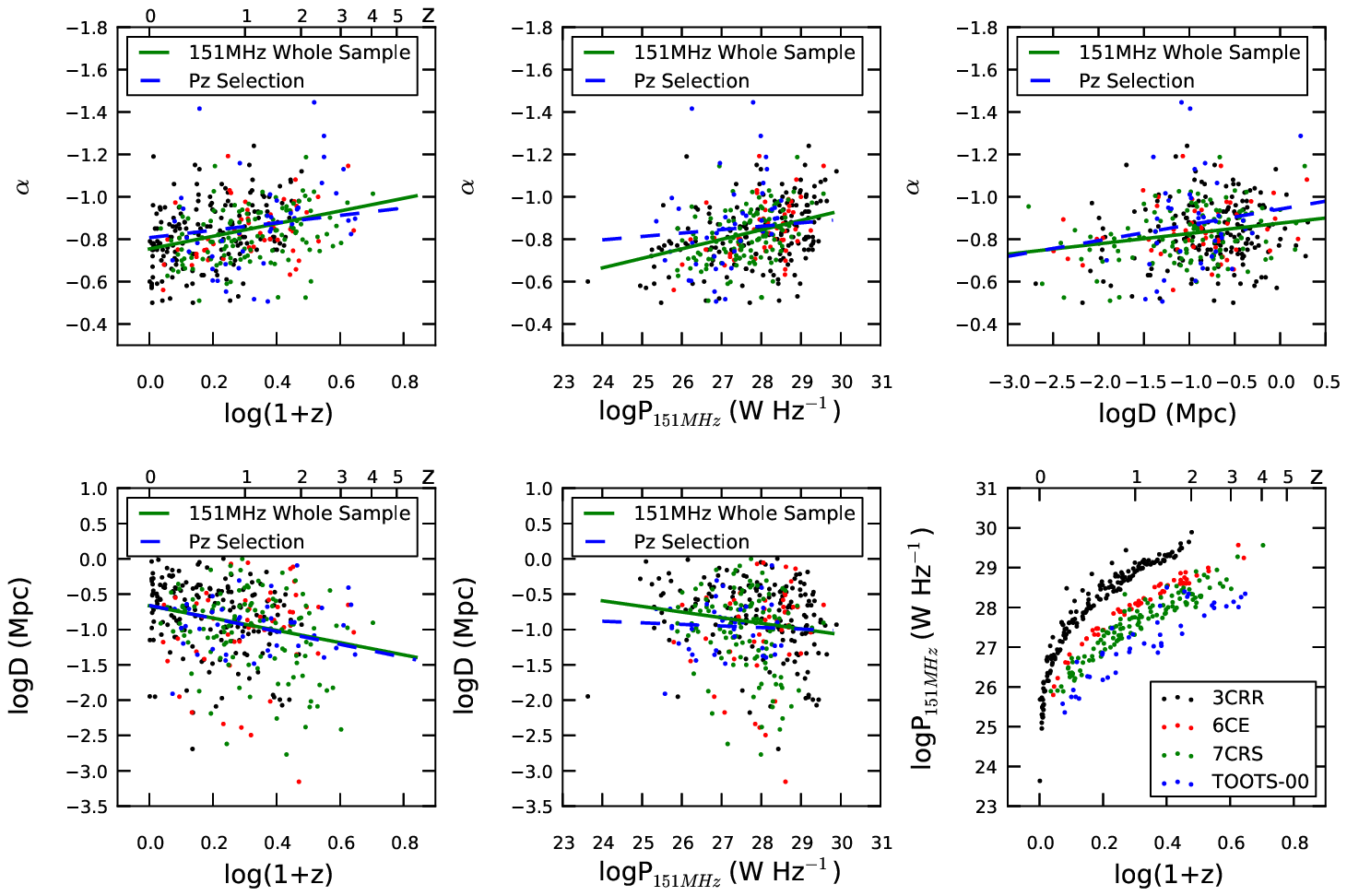}
		\vspace{-2mm}
		\includegraphics[width=\textwidth]{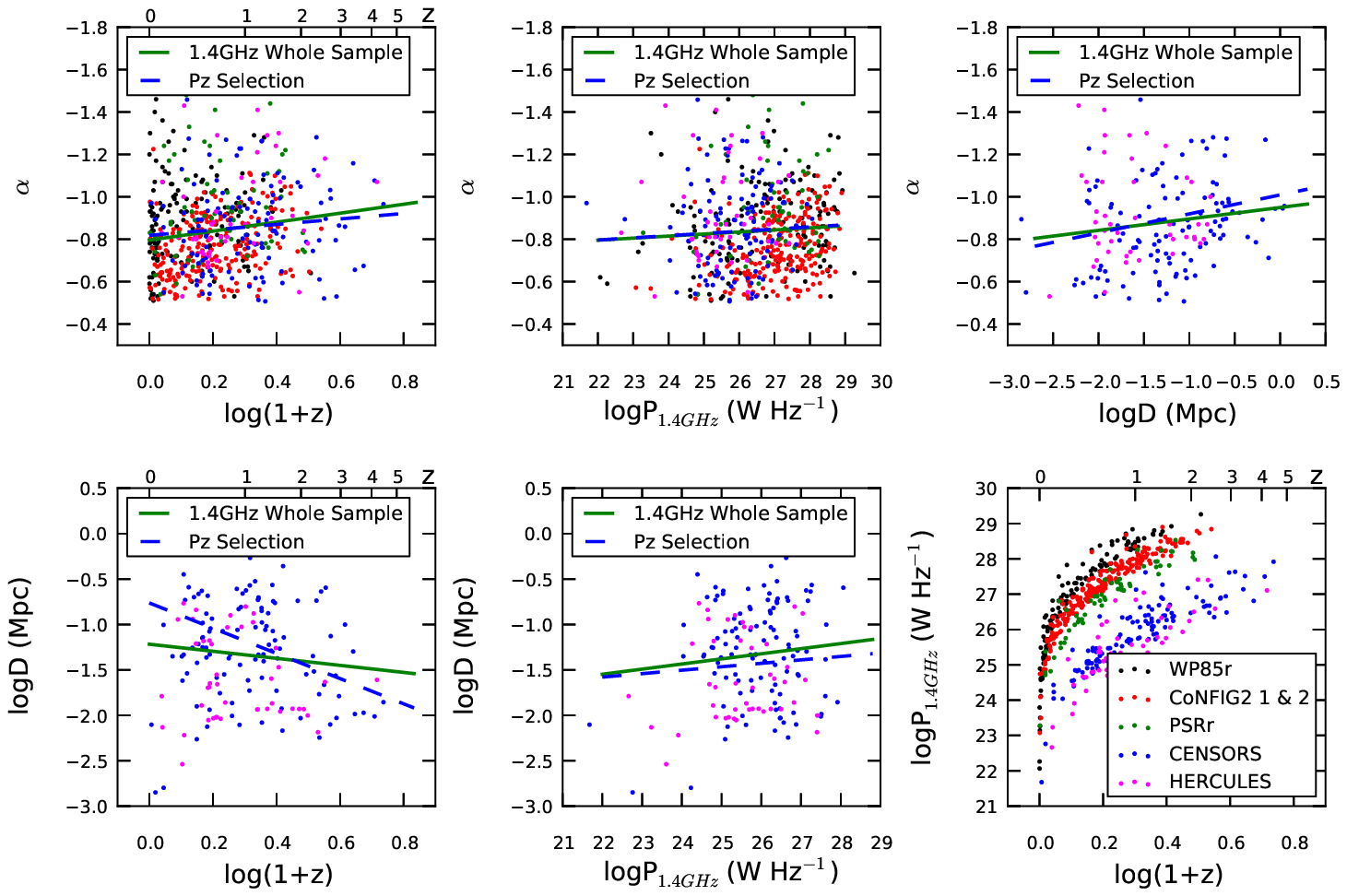}
\caption{\label{Fig3} \small The z$\alpha$PD planes for the 151 MHz (upper 6 panels) and 1.4 GHz (lower 6 panels) selected samples. Spectral indices are measured between 1400-151 MHz for 6CE, 7CRS, TOOTS-00 and CoNFIG1\&2r, 750-151 MHz for 3CRR, 5-2.7 GHz for WP85r, 2-1.4 GHz for PSRr, 1400-325 MHz for CENSORS and between 1400 MHz and 610 MHz for Hercules. Only sources with $\alpha$ steeper than $-$0.5 are utilised. The solid green lines indicate the best fitting straight line to the data. The blue dashed lines indicate the linear fit repeated for a Malmquist-bias-free section of the P-z plane, as defined in Figure \ref{Fig1}. Note that only CENSORS and Hercules are included in the linear size figures for the 1.4 GHz samples (see text).}
\end{center}
\end{figure*}

\setcounter{table}{1}
\begin{table*}
\caption{\label{Tab2} Spearman Rank Correlation Coefficients and the associated 2-tailed p-value for various combinations of PD$\alpha$z. A * denotes that only the CENSORS and Hercules samples were used in measuring the correlation, as these are the only two high frequency selected samples with size information readily available.}
\begin{tabular}{|l|l|l|l|l|}
\hline
  \multicolumn{1}{|c|}{Combination} &
  \multicolumn{1}{|c|}{151 MHz(All)} &
  \multicolumn{1}{c|}{151 MHz(P-z)} &
  \multicolumn{1}{c|}{1.4 GHz(All)} &
  \multicolumn{1}{c|}{1.4 GHz(P-z)} \\
\hline
\hline
logP-log(1+z)&r=\phantom{$-$}0.72, p=0.000 &r=\phantom{$-$}0.02, p=0.780 &r\phantom{$*$}=\phantom{$-$}0.61, p=0.000 &r\phantom{$*$}=$-$0.19, p=0.030 \\
logP-logD&r=$-$0.14, p=0.010 &r=\phantom{$-$}0.07, p=0.350 &r*=\phantom{$-$}0.01, p=0.860 &r*=\phantom{$-$}0.01, p=0.920 \\
logP-$\alpha$&r=$-$0.36, p=0.000 &r=$-$0.10, p=0.180 &r\phantom{$*$}=$-$0.11, p=0.010 &r\phantom{$*$}=\phantom{$-$}0.01, p=0.920 \\
log(1+z)-$\alpha$&r=$-$0.34, p=0.000 &r=$-$0.15, p=0.050 &r\phantom{$*$}=$-$0.19, p=0.000 &r\phantom{$*$}=$-$0.14, p=0.100 \\
log(1+z)-logD&r=$-$0.25, p=0.000 &r=$-$0.15, p=0.050 &r*=$-$0.10, p=0.211 &r*=$-$0.53, p=0.000 \\
logD-$\alpha$&r=$-$0.16, p=0.001 &r=$-$0.26, p=0.000 &r*=$-$0.18, p=0.028 &r*=$-$0.25, p=0.080 \\
\hline
\end{tabular}
\end{table*}

\section{Observable Trends}

The complete samples detailed previously provide excellent coverage of the Pz$\alpha$D parameter space. In Figure \ref{Fig3}, the logP, logD, $\alpha$ and log(1+z) planes are plotted, along with best fitting straight lines to the data. By eye, the data appear to display similar dependencies of spectral index to those reported by \citet{BlundellAstron.J.117:677-7061999} for the 3CRR, 6CE and 7CRS combined complete radio samples, namely that observed spectral indices steepen with linear size, redshift and radio power (upper panels). Equations of the linear dependencies fitted for spectral index on luminosity, linear size and redshift are given in Tables 5 and 6.

As can be clearly seen from the P-z panels in Figure \ref{Fig3}, the use of only one complete sample means that a strong, dominating, P-z correlation due to Malmquist bias is present, and this makes disentangling the various dependancies between P, z, $\alpha$ and D very difficult. The addition of several complete samples mitigates this correlation somewhat, and indeed many previous studies, e.g. \citet{BlundellAstron.J.117:677-7061999} argue that the combination of several complete samples essentially removes the Malmquist bias.

With the excellent coverage of the P-z plane afforded by our nine complete samples, we are able to test if this is indeed the case. We select a Malmquist-bias-free section of the P-z plane for both the high and low frequency selected samples, covering a large range in both redshift and radio power (see Figure \ref{Fig1} for the area definition), and repeat the linear fits (plotted as blue dashed lines in Figure \ref{Fig3}). This utilises 186 sources in 151 MHz samples, and 133 sources in the 1.4 GHz samples (reduced to 56 sources when investigating linear size, as for the 1.4 GHz samples only CENSORS and Hercules have readily available size information). The D-$\alpha$ relation appears to increase in strength when the residual P-z correlation is removed, whilst the z-$\alpha$ and P-$\alpha$ decrease in strength. 

In Table \ref{Tab2} we present the Spearman rank correlation coefficients for the relations plotted in Figure \ref{Fig3}. Also listed is the 2-tailed p-value, which gives an approximate indication of the probability of an uncorrelated system having a Spearman correlation at least as strong as the one calculated from the observed data. The table illustrates several important points. Firstly it shows that the P-z correlation dominates, even when several complete samples are coadded and analysed together, in other words, {\em simply adding several complete samples does not provide sufficient coverage of the P-z plane to fully remove the dominant P-z correlation}. Secondly it reveals that the D-$\alpha$, D-z and z-$\alpha$ correlations are the strongest observed in both high and low frequency selected samples. The z-$\alpha$ correlation is stronger than the P-$\alpha$ correlation (correlations between P, $\alpha$ and D more or less disappear once the P-z correlation is removed). Of particular relevance to this study is the fact that the variation of observed $\alpha$ with size and redshift is relatively weak for both samples.

\subsection{Principle Component Analysis}

Analysing fully covered sections of the P-z plane has shown that relations between P, z, $\alpha$ and D are strongly coupled to the Malmquist bias. With this in mind, we utilise another statistical test, Principle Component Analysis (PCA), which is a technique designed to pick out the intrinsic, dominant linear correlations existing in a multi-variable dataset, as opposed to secondary correlations arising due to combinations of others (in this case, particularly the Malmquist bias). The method of PCA involves calculating the eigenvalues and eigenvectors, composed of linear combinations of the normalised input parameters, which span the directions of maximum variance in the input dataset. These eigenvectors and eigenvalues describe the intrinsic correlations present in the dataset (principal components), along with the percentage of the variance in the data that each explain.The results of PCA are most commonly presented in table form, listing each of the principal components, the percentage of
the data variance that they explain, and the composition of each principle component. Each principle component is composed
of a normalised combination of the entered variables, in this case $\alpha$, log(1+z), logP and logD, as PC = x$_1$$\alpha$ +x$_2$log(1+z) + x$_3$logP + x$_4$logD, and the final four columns in the table present x$_1$, x$_2$, x$_3$,and x$_4$, showing the relative contributions of each variable for each principle component.

We first look at a low frequency selected sample, composed of the 3CRR, 6CE, 7CRS and TOOTS samples, and perform a PCA analysis (see Table \ref{Tab3}, upper). The P-z correlation dominates (i.e. the first principle component is along an axis primarily composed of P and z), contributing roughly half of the observed variance. A further $\sim$30\% variance is contributed by a D-$\alpha$ anticorrelation, whereby sources of larger size have steeper spectra. The final two components largely just account for scatter around the two dominating independent relations between P-z and D-$\alpha$. This is an important finding, which is consistent with that demonstrated by the previous section, that the P-z correlation remains dominant even when a large collection of complete samples is used. 

Although PCA should successfully identify all underlying independent correlations in the data, we ran the analysis again on just the subsamples in a well-covered region of the P-z plane, thereby removing the selection effect (see Figure \ref{Fig1}). A second motivation for doing this is to restrict analysis to only high-power radio sources, thus studying a relatively uniform population (extended double radio sources), with little contamination from low power sources which are often unresolved (see for example, \citealt{Baldi2009}). In this case, the observed variance can be attributed to two independent relations, each giving an almost equal contribution to the variance (see Table \ref{Tab3}, lower). The largest contributer, at 33\%, is an anticorrelation between $\alpha$ and D as found for the whole sample, followed by 28\% contribution between D and z. The third 24\% contribution arises almost solely along the logP axis, uncorrelated with the other parameters. This confirms our earlier findings, that  D-$\alpha$ and D-z relations are intrinsic to the dataset, irrespective of the presence of Malmquist bias. 

The results for a high frequency selected collection of samples (composed of CENSORS and Hercules) show broadly similar results, albeit with some difference in the detail (see Table \ref{Tab4}). For the entire sample, the results are very similar, again with approximately half the variance being acounted for by the P-z correlation, and a further 30\% by a D-$\alpha$ correlation. The main difference is that this latter correlation is weakened somewhat by the third component. If we then select a well covered section of the P-z plane (see Figure \ref{Fig1}), then similarly to the low frequency data, approximately 40\% of the variance is accounted for by a D-z anticorrelation, followed by a 32\% $\alpha$-D, P anticorrelation (see Table \ref{Tab4}, lower). A weaker P-$\alpha$ correlation accounts for the large remainder of the variance, which removes the weak P-$\alpha$ anticorrelation contribution of the second component. 

From this analysis we can tentatively conclude that there are two firm independent relations present in both datasets, between D and $\alpha$, and between P and z. By utilising the full P-z coverage subsamples, we confirm that the D-$\alpha$ correlation is fully independent of Malmquist bias. In the subsamples restricted to be high power sources with good P-z coverage, a strong D-z anticorrelation is also seen. That the D-$\alpha$ correlation is slightly stronger in the low frequency dataset, and the D-z anticorrelation stronger in the high frequency sample is most likely down to the differing types of sources which low and high frequency selected samples collect. Low frequency samples will primarily be composed of lobe dominated sources, suffering little in the way of orientation bias, and hence a large proportion of large, steep spectrum sources. High frequency selected samples will include many more beamed, core dominated and young GPS/CSS sources, and less classical lobed dominated sources. 

The correlation between D and $\alpha$ arises due to aging of the radio sources. As a source gets older, it increases in size and the spectrum steepens with age. The physical cause of the anticorrelation between D and z is subject to more debate. It could arise as the result of the environment at high redshift, or as a result of sources at high redshift being more likely to be younger, and hence smaller (cf. \citealt{Blundell1999}). It is interesting to note that despite a Spearman rank test (see Table \ref{Tab2}) suggesting the presence of a correlation between z and $\alpha$ almost as strong as that between D and z, the Principle Component Analysis does not clearly identify an independent z-$\alpha$ relation in either the high or low frequency selected samples, suggesting that the correlation observed may be largely a result of selection effects. One which may be present is that between radio power and linear size. It is thought that radio sources follow ‘tracks’ on the P-D plane, beginning with high power, small sources, and evolving into lower power, larger sources in time (see for example \citealt{Kaiser2007}). Individual low frequency samples, which are more sensitive to extended radio lobes, show a trend for radio power to increase as linear size decreases, which could arise from the combination of Malmquist bias and the D-z correlation (and indeed, this correlation weakens substantially once Malmquist bias is removed). In a collection of low frequency selected samples, this trend in conjunction with any remaining Malmquist bias and the D-$\alpha$ correlation would naturally lead to an extrinsic contribution to the z-$\alpha$ correlation.
\setcounter{table}{2}
\begin{table}
\begin{tabular}{|l|l|l|l|l|l|}
\hline
  \multicolumn{1}{|c|}{PC} &
  \multicolumn{1}{|c|}{\%} &
  \multicolumn{1}{c|}{$\alpha_{obs}$} &
  \multicolumn{1}{c|}{log(1+z)} &
  \multicolumn{1}{c|}{logP$_{151 MHz}$} &
  \multicolumn{1}{c|}{logD(Mpc)} \\
\hline
\hline
1&49&\phantom{$-$}0.38&$-$0.65&$-$0.64&\phantom{$-$}0.19\\
2&29&$-$0.60&$-$0.10&$-$0.02&\phantom{$-$}0.79\\
3&15&\phantom{$-$}0.70&\phantom{$-$}0.20&\phantom{$-$}0.38&\phantom{$-$}0.57\\
4&7.0&\phantom{$-$}0.06&\phantom{$-$}0.72&$-$0.67&\phantom{$-$}0.12\\
\hline
\end{tabular}

\begin{tabular}{|l|l|l|l|l|l|}
\hline
  \multicolumn{1}{|c|}{PC} &
  \multicolumn{1}{|c|}{\%} &
  \multicolumn{1}{c|}{$\alpha_{obs}$} &
  \multicolumn{1}{c|}{log(1+z)} &
  \multicolumn{1}{c|}{logP$_{151 MHz}$} &
  \multicolumn{1}{c|}{logD(Mpc)} \\
\hline
\hline
1&33&\phantom{$-$}0.71&$-$0.15&$-$0.30&$-$0.62\\
2&28&$-$0.20&\phantom{$-$}0.85&\phantom{$-$}0.10&$-$0.48\\
3&24&\phantom{$-$}0.25&$-$0.13&\phantom{$-$}0.95&$-$0.15\\
4&15&$-$0.63&$-$0.48&\phantom{$-$}0.003&$-$0.61\\
\hline
\end{tabular}
\caption{\label{Tab3} Upper Table: Principle Component Analysis for the low frequency selected samples, comprising 375 sources with $\alpha$ $<$ $-$0.5. Lower Table: The same analysis repeated for a well covered section of the P-z plane: logP(W Hz$^{-1}$)=27.75-29.5 z=0.5-3.5, using 186 sources.}
\end{table}

\setcounter{table}{3}
\begin{table}
\begin{tabular}{|l|l|l|l|l|l|}
\hline
  \multicolumn{1}{|c|}{PC} &
  \multicolumn{1}{|c|}{\%} &
  \multicolumn{1}{c|}{$\alpha_{obs}$} &
  \multicolumn{1}{c|}{log(1+z)} &
  \multicolumn{1}{c|}{logP$_{1.4 GHz}$} &
  \multicolumn{1}{c|}{logD(Mpc)} \\
\hline
\hline
1&47&$-$0.12&\phantom{$-$}0.70&\phantom{$-$}0.70&\phantom{$-$}0.01\\
2&29&\phantom{$-$}0.66&\phantom{$-$}0.12&\phantom{$-$}0.02&$-$0.74\\
3&22&$-$0.74&\phantom{$-$}0.08&$-$0.19&$-$0.64\\
4&3.0&$-$0.09&$-$0.70&\phantom{$-$}0.69&$-$0.18\\
\hline
\end{tabular}

\begin{tabular}{|l|l|l|l|l|l|}
\hline
  \multicolumn{1}{|c|}{PC} &
  \multicolumn{1}{|c|}{\%} &
  \multicolumn{1}{c|}{$\alpha_{obs}$} &
  \multicolumn{1}{c|}{log(1+z)} &
  \multicolumn{1}{c|}{logP$_{1.4 GHz}$} &
  \multicolumn{1}{c|}{logD(Mpc)} \\
\hline
\hline
1&42&\phantom{$-$}0.005&\phantom{$-$}0.71&\phantom{$-$}0.48&$-$0.52\\
2&32&\phantom{$-$}0.71&$-$0.08&$-$0.46&$-$0.52\\
3&19&$-$0.67&\phantom{$-$}0.15&$-$0.62&$-$0.39\\
4&7.0&$-$0.21&$-$0.69&\phantom{$-$}0.42&$-$0.56\\
\hline
\end{tabular}
\caption{\label{Tab4}Upper Table: Principle Components Analysis on the GHz frequency selected samples of CENSORS and Hercules, comprising 158 sources with $\alpha$ $<$ $-$0.5. Lower Table: Repeated for a well covered selection of the P-z plane: logP(W Hz$^{-1}$)=26.25-29 z=1-4.5, using 56 sources.}
\end{table}

\setcounter{table}{4}
\begin{table*}
\caption{\label{Tab5}The results of fitting functions of $\alpha$ dependent on z, P and D for the low frequency selected samples. We assume a measurement error of 0.1 in $\alpha$ for all fits, for the determination of the reduced $\chi^2$ and $\sigma$.}
\begin{tabular}{|l|l|l|l|l|l|l|l|}
\hline
  \multicolumn{1}{|c|}{Model} &
  \multicolumn{1}{|c|}{Sample} &
  \multicolumn{1}{c|}{r$\chi^2$} &
  \multicolumn{1}{c|}{$\sigma$} &
  \multicolumn{1}{c|}{a$_1$} &
  \multicolumn{1}{c|}{a$_2$} &
  \multicolumn{1}{c|}{a$_3$} &
  \multicolumn{1}{c|}{a$_4$} \\

\hline
\hline
$\alpha$=-0.8&whole&2.47&0.15&-&-&-&-\\
 &Pz&2.53&0.14&-&-&-&-\\
$\alpha$=a$_1$log(1+z)+a$_2$&whole&1.98&0.15&-0.30(0.03)&-0.75(0.01)&-&-\\
 &Pz&1.01&0.14&-0.17(0.06)&-0.80(0.02)&-&-\\
$\alpha$=a$_1$logP+a$_2$&whole&1.99&0.15&-0.04(0.01)&0.40(0.14)&-&-\\
 &Pz&1.02&0.14&-0.02(0.02)&-0.41(0.43)\\
$\alpha$=a$_1$logD+a$_2$&whole&2.12&0.15&-0.05(0.01)&-0.88(0.01)&-&-\\
 &Pz&0.92&0.14&-0.07(0.01)&-0.94(0.01)\\
$\alpha$=a$_1$log(1+z)+a$_2$logP+a$_3$&whole&1.95&0.15&-0.17(0.05)&-0.03(0.01)&-0.09(0.18)&-\\
 &Pz&1.00&0.14&-0.17(0.06)&-0.01(0.02)&-0.59(0.44)&-\\
$\alpha$=a$_1$logP+a$_2$logD+a$_3$&whole&1.87&0.14&-0.05(0.01)&-0.06(0.01)&0.49(0.14)&-\\
 &Pz&0.92&0.14&-0.02(0.02)&-0.07(0.01)&-0.44(0.44)&-\\
$\alpha$=a$_1$log(1+z)+a$_2$logD+a$_3$&whole&1.83&0.14&-0.36(0.03)&-0.07(0.01)&-0.80(0.01)&-\\
 &Pz&0.88&0.13&-0.25(0.06)&-0.08(0.01)&-0.86(0.02)&-\\
$\alpha$=a$_1$log(1+z)+a$_2$logP+a$_3$logD+a$_4$&whole&1.80&0.14&-0.25(0.05)&-0.02(0.01)&-0.07(0.01)&-0.18(0.19)\\
 &Pz&0.88&0.13&-0.25(0.06)&-0.01(0.02)&-0.08(0.01)&-0.70(0.44)\\
\hline
\end{tabular}
\end{table*}

\setcounter{table}{5}
\begin{table*}
\vspace{1cm}
\caption{\label{Tab6}The results of fitting functions of $\alpha$ dependent on z, P and D for the high frequency selected samples. We assume a measurement error of 0.1 in $\alpha$ for all fits, for the determination of the reduced $\chi^2$ and $\sigma$. Models marked with a * use only CENSORS and Hercules, the only two high frequency selected samples for which there is size data readily available.}
\begin{tabular}{|l|l|l|l|l|l|l|l|}
\hline
  \multicolumn{1}{|c|}{Model} &
  \multicolumn{1}{|c|}{Sample}&
  \multicolumn{1}{c|}{r$\chi^2$} &
  \multicolumn{1}{c|}{$\sigma$} &
  \multicolumn{1}{c|}{a$_1$} &
  \multicolumn{1}{c|}{a$_2$} &
  \multicolumn{1}{c|}{a$_3$} &
  \multicolumn{1}{c|}{a$_4$} \\
\hline
\hline
$\alpha$=-0.8&whole&4.10&0.20&-&-&-&-\\
 &Pz&3.70&0.19&-&-&-&-\\
$\alpha$=a$_1$log(1+z)+a$_2$&whole& 2.75 & 0.20 & -0.21(0.03) & -0.80(0.01) & -  & - \\
 &Pz&1.56 & 0.18 & -0.13(0.04) & -0.81(0.01) & - &- \\
$\alpha$=a$_1$logP+a$_2$ &whole& 2.81 & 0.20 & -0.01(0.003) & -0.58(0.09) & -  & - \\
 &Pz&1.57 & 0.19 & -0.01(0.01) & -0.57(0.20) & - &- \\
$\alpha$*=a$_1$logD+a$_2$ &whole& 3.20 & 0.21 & -0.054(0.01) & -0.95(0.02) & -  & - \\
 &Pz&0.81 & 0.18 & -0.09(0.02) & -1.01(0.04) & - &- \\
$\alpha$=a$_1$log(1+z)+a$_2$logP+a$_3$&whole&2.74&0.20&-0.24(0.03)&0.006(0.003)&-0.95(0.10)&-\\
 &Pz&1.56&0.18&-0.13(0.04)&-0.0001(0.01)&-0.81(0.22)&-\\
$\alpha$*=a$_1$logP+a$_2$logD+a$_3$&whole&3.20&0.21&-0.004(0.01)&-0.05(0.01)&-0.85(0.20)&-\\
 &Pz&0.79&0.18&-0.06(0.03)&-0.09(0.03)&0.58(0.80)&-\\
$\alpha$*=a$_1$log(1+z)+a$_2$logD+a$_3$&whole&3.14&0.21&-0.18(0.05)&-0.06(0.01)&-0.90(0.02)&-\\
 &Pz&0.75&0.17&-0.41(0.11)&-0.12(0.01)&-0.90(0.02)&-\\
$\alpha$*=a$_1$log(1+z)+a$_2$logP+a$_3$logD+a$_4$&whole&3.02&0.20&-0.63(0.1)&0.08(0.02)&-0.09(0.01)&-2.90(0.38)\\
 &Pz&0.74&0.17&-0.42(0.13)&0.01(0.04)&-0.12(0.03)&-0.99(0.94)\\
\hline
\end{tabular}
\end{table*}

\begin{figure*}
\begin{center}
                \includegraphics[width=\textwidth]{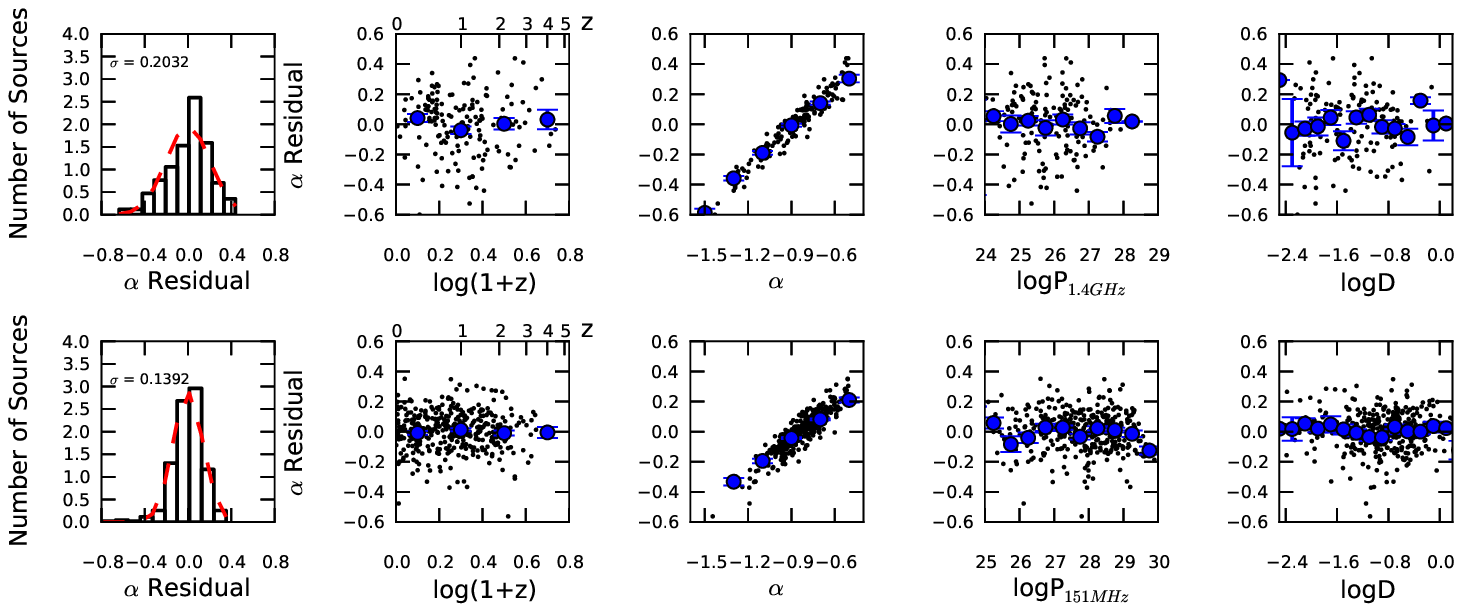}
\caption{\small \label{Fig4}Fitting the function $\alpha$=a$_1$log(1+z) + a$_2$logP + a$_3$logD + a$_4$ to the CENSORS and Hercules combined samples (upper five panels), and to the 3CRR, 6CE, 7CRS and TOOTS combined samples (lower five panels).The panels on the extreme left show the distributions of the spectral index $\alpha$ residuals (observed $\alpha$ minus the model predicted $\alpha$), and the next four panels on each line show how the $\alpha$ residual depends on z, $\alpha$, P, and D. Small points are the raw data, large points are the binned means. Fitting the functions clearly removes trends in z,D, and P but large scatter remains, as indicated in the $\alpha$ plot.}
\end{center}
\end{figure*}

\section{A Large Intrinsic Scatter in $\alpha$}

Given the independent trends between spectral index, linear size and redshift, identified both visually and by the PCA analysis, an attempt was made to fit an analytical form to the spectral index using linear size, luminosity and redshift.  Again, as detailed in Section 3, we use only sources with an observed spectral index steeper than $-$0.5.

Table \ref{Tab5} and Table \ref{Tab6} list the best fitting coefficients for each relation modeled. We began with very simple linear fits, and progressed to fitting planes modeling all four variables.  We can see clearly that both the reduced $\chi^2$ and the residual standard deviation decrease, albeit by a small amount, with the inclusion of additional variables in the model for both the high and low frequency selected samples. A plane fit of all four variables gives the best fit, and the smallest deviation in $\alpha$ residuals for both low and high frequency selected samples. The best fitting model is illustrated in Figure \ref{Fig4}. Although the plane model manages to successfully remove the trends between spectral index and linear size, radio power and redshift, the key finding is that it is unable to predict the observed $\alpha$. The intrinsic scatter in $\alpha$ is much greater than that arising from any physical trends with other observables present in the datasets.

Whilst this was a simplistic approach designed to see if it was possible to predict the observed spectral index with any success from other properties, it should be noted that much more complex models, incorporating the physics of radio sources can reproduce the observed luminosity, linear size and redshift distributions with some success, but struggle to reproduce $\alpha$ (see for example \citealt{BaraiAstrophys.J.658:217-2312007}).

It is very clear that the correlations between $\alpha$ and size, luminosity and redshift are weak. The results of this suggest that the use of spectral index alone is unlikely to be efficient in selecting high redshift radio sources. The equally strong D-z correlation indicates that inclusion of radio size information may increase the efficiency of selection based solely on radio observables.
\begin{figure}
\begin{center}
		\includegraphics[width=\columnwidth]{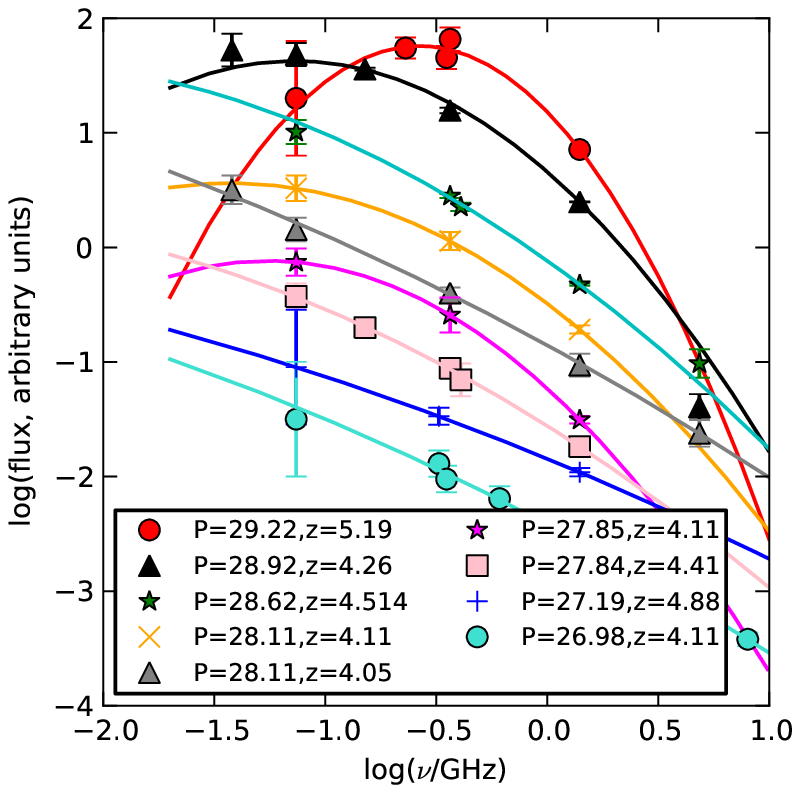}        
\caption{\small \label{Fig5}Radio spectra for the nine highest redshift z$>$4 radio galaxies known. Fluxes were obtained from the NASA Extragalactic Database (NED), at frequencies ranging from 38 MHz to 8 GHz. The flux scale is offset by a small arbitrary amount for each source to allow the shapes of the radio spectra to be compared.  TN J0924 (z=5.19), J1639 (z=4.88) and CEN69 (z=4.11) have spectral indices which imply that they should be detected in the VLSS (\protect \citealt{Cohen2007}) assuming a straight spectrum, however all three are not. We measure the noise ($\sigma$) at each source position in the VLSS maps, and take the 2$\sigma$ value as the upper limit 74 MHz flux, and assume a one $\sigma$ error. Note that interestingly, the two highest redshift known sources are easily detectable at 1.4 GHz in the NVSS, at 71.5 mJy and 21.8 mJy respectively, but would not be detected in any currently existing 150 MHz or 74 MHz surveys. The vast majority of the spectra are classified as compact steep spectrum, most flattening towards lower frequencies (four of these are potentially peaked). Only one (7C1814, at the lowest redshift) is confirmed as straight over the frequency range 74 MHz to 5 GHz. This negates the common assumption that high redshift, USS sources display no curvature over a large frequency range. The sources are ordered in P$_{1.4 GHz}$ (calculated with full curvature information), and it can be clearly seen that the most powerful sources are more likely to display significant curvature. It is also worth noting that all but one of these sources (7C1814) are compact, and have  $\theta$$<$6''. }
\end{center}
\end{figure}
\section{The Origin of the $\alpha$-z Correlation in Flux Limited Samples}

The tendancy for observed spectral indices to steepen with redshift has been attributed to a k-correction, where as the source spectrum is redshifted, a steeper part of the spectrum is sampled. How much of an effect this is has been a source of much debate in the literature. It has also been suggested that the strength of the z-$\alpha$ correlation increases with frequency, as high frequency parts of the radio spectrum undergo more significant synchrotron losses.

\citet{KlamerMon.Not.Roy.Astron.Soc.371:852-8662006} find the majority of their USS sample display no curvature, and also cite the well studied high redshift source 4C41.17, at z=3.8, as having a straight radio spectrum from 26 MHz to GHz frequencies. They therefore infer that the k-correction is irrelevant for high-z USS sources. However, based on this, we cannot simply conclude that the contribution to any z-$\alpha$ correlation from the k-correction is negligible. In fact, Figure \ref{Fig5} shows the radio spectra of all currently known z$>$4 radio galaxies - the majority of which show some evidence of curvature in the observed radio spectrum. Most of the currently known radio galaxies at z$>$4, display a compact steep radio spectrum, with curvature occurring at low observed frequencies ($\sim$100 MHz), data which Klamer et al did not have for their sample. \citet{Bornancini2007} also confirm the presence of curvature at low MHz frequencies for their USS sample.

To quantify the potential effect of the k-corrections, we used two samples, CoNFIG and 7CRS. These two samples have the best multi-frequency coverage, and hence most accurately determined radio spectra. Rest frame spectral indices are calculated from fitting a 2nd-order polynomial (logS$_\nu$=a$_1$+a$_2$log$\nu$+a$_3$log$^{2}$$\nu$) to the radio spectrum for each source, and measuring the gradient ($\alpha$=a$_2$+2a$_3$log$\nu$) at the desired frequency; details of this will be presented in Ker et al. (in prep). A 2nd order polynomial fit provides a good fit to the radio spectra of the vast majority of sources in each sample. 

\begin{figure}
\begin{center}
                \includegraphics[width=\columnwidth]{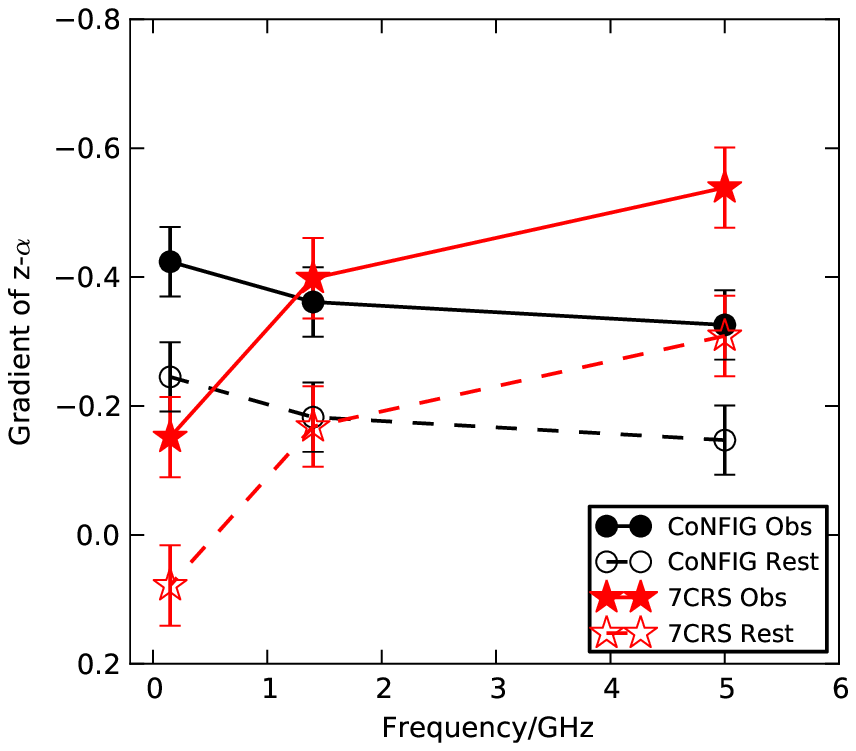}
		\includegraphics[width=\columnwidth]{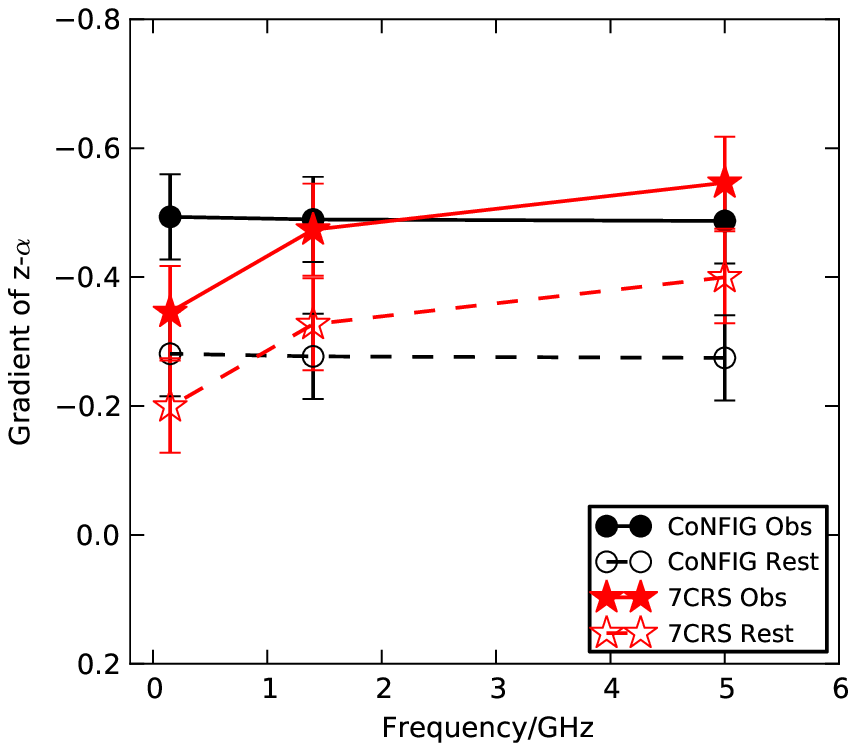}
\caption{\small \label{Fig6}The upper panel shows the gradient `a' from fitting $\alpha$=a*log(1+z)+b to rest-frame and observed spectral indices for CoNFIG and 7CRS. Whilst both samples show a clear decrease in the gradient of the z-$\alpha$ correlation when k-corrections are applied, only 7CRS shows a marked increase in strength with the frequency at which $\alpha$ is measured. The lower figure shows the same fit again for CoNFIG and 7CRS, but this time with all known quasars and compact objects (classifed as compact in CoNFIG, or less than 30kpc in size in 7CRS) removed. The gradients become very similar at all frequencies and for both samples when only extended sources are considered. These figures clearly show that the k-correction can be responsible for up to 50\% of the z-$\alpha$ gradient observed in flux limited radio samples.}
\end{center}
\end{figure}

We then performed a simple linear fit to the observed and rest-frame spectral index measured at three frequencies as a function of log(1+z) for both CoNFIG and 7CRS (see Figure \ref{Fig6}). We performed the fit only on sources with an observed spectral index between 1.4 GHz and 151 MHz less than $-$0.5 and with a well determined radio spectrum. The gradients of these fits then reflect the strength of the z-$\alpha$ correlation present (if any). The results we obtain are striking. For 7CRS we confirm that the gradient of both the observed and the rest-frame z-$\alpha$ correlation increases with the frequency at which $\alpha$ is measured, as first reported by \citet{BlundellAstron.J.117:677-7061999}. We also see that the measured z-$\alpha$ correlation is approximately twice as steep in the observed-frame than in the rest-frame (dependent on frequency). It is also worthwhile noting that for 7CRS, contrary to the z-$\alpha$ correlation, the D-$\alpha$ correlation strengthens in the rest-frame.

Similarly, for CoNFIG we see that the observed-frame correlation can be 50\% steeper, or more at all frequencies than that measured in the rest-frame. However the increase in gradient with frequency is not seen. We suggest that this is because GHz selected samples pick out very different proportions of various types of radio source, favouring young GPS/CSS, core and beamed sources (much higher orientation bias). 

To test this, we ran the fits again, this time excluding all known quasars, and objects classified as compact in CoNFIG, and sources with a size less than 30kpc in 7CRS. This ensures that the vast majority of sources included in both samples will be lobe dominated and working against the IGM, and are not heavily contaminated by beamed sources or are sources so small that they are still propagating through the medium of their host galaxy rather than the IGM. The results are plotted in the bottom panel of Figure \ref{Fig6}, and the difference is clear to see. Both CoNFIG and 7CRS now follow very similar relations, both displaying observed gradients which are approximately twice as strong as the rest-frame gradients, but which are now largely independent of the frequency at which $\alpha$ is measured. It is interesting to note that the strength of the gradient for observed $\alpha$ for both samples is very similar to that determined by Ubachukwu et al. (1995) for a sample of radio galaxies compiled from the 3CRR and WP85 samples, again excluding compact sources.   

Our results confirm that once the k-correction is removed, a weak correlation between z and $\alpha$ remains for extended radio galaxies, which would fit in with a scenario where lobes are working against a denser environment at higher redshift, and hence high frequency losses are greater. \citet{Miley2008} note, however, that it is very difficult to reproduce the observed z-$\alpha$ relation from this somewhat simplistic density-dependent effect. They suggest that as the density of gas around high redshift sources has been observed to be highly inhomogeneous, and denser close to the nucleus of the galaxy, that the ultra-steep radio spectra are produced by some as yet unknown mechanism within the host galaxy, rather than by the IGM conditions through which the radio lobes are propagating.

We also conclude that GHz selected samples have a much greater orientation bias present, which can disguise the presence of the z-$\alpha$ correlation displayed by extended radio galaxies. We have also successfully demonstrated that the k-correction is not negligible when measuring the strength of any z-$\alpha$ correlation, and can be responsible for more than 50\% of the strength of the observed gradient in a flux-limited sample.

\section{Implications for High Redshift Searches}

The data collected for the nine complete samples allows us to measure for the first time, the efficiency of the three most commonly used methods in the literature for searching for high redshift radio galaxies, namely radio spectral index, angular size and \emph{K}-band magnitude. We are looking for a set of criteria which minimises the size of the selected subsamples that would require follow up observations, whilst maximising the number of high-z galaxies retained in this sample. We assume a definition of highest efficiency as maximising the difference between the total fraction of the sample recovered, and the total fraction of high-z sources recovered, with each increasing cut in the selection parameter under study. We choose to consider `z$>$2' radio galaxies as high-z sources, as for the datasets under consideration this provides the optimal compromise between maximising the redshift whilst still maintaining sufficient high-z sources to allow a robust analysis. For comparison, we also show the analysis repeated for z$>$3 where possible, albeit with much lower number statistics (we have 10 z$>$3 radio galaxies with spectroscopic redshifts - see Table \ref{Tab7} - and 6 with photometric redshifts in our samples). As there are only approximately 50 radio galaxies with z$>$3 known (\citealt{Ishwara-Chandra2010}), our samples are hence representative of the highest known redshift radio galaxy parameter space).

\subsection{Spectral Index Selection}
As discussed above, an initial steep radio spectral index cut is an extremely popular method of reducing very large radio samples down to managable sizes for imaging and spectroscopic follow-up, in order to locate high redshift sources. We now investigate whether a first spectral index cut does indeed recover a significant proportion of high-z sources present in the samples. Many recent studies in the literature base searches for high-z radio galaxies on the assumption that they may be distinguished by a steep spectrum. \citet{Ishwara-Chandra2010} provide a list of the highest redshift, z$>$3, known radio galaxies, 47 in number, the vast majority of which have been selected from an USS sample. However \citet{Jarvis2009}, also recently reported the discovery of the second highest redshift radio galaxy known, a source which they noted clearly does not have a ultra steep spectral index (see Figure \ref{Fig5}). Work with the DRaGONS study (\citealt{Schmidt2006}), which uses a large, bright radio sample from the 1.4 GHz FIRST survey with redshifts estimated from the K-z relation, also suggests that even with a relatively flat spectral index selection criterion of $\alpha$ $<$ $-$0.8, one third of the z$>$2 sources are missed. In Table \ref{Tab7}, we present a list of the 10 radio galaxies with a confirmed spectroscopic redshift of z$>$3 from all of the samples used in this study. Five of these do have a steep $\alpha$ $<$ $-$1.0, however the remainder display a wide variety of spectral indices. 

In studies utilising an ultra steep selection criterion, an often used argument to justify the use of steep spectral index cut-offs is the apparent strong shift in redshift distribution to high redshift. However, in the majority of these studies (e.g. \citealt{Breuck2000}, \citealt{Bryant2009}), the samples are very large (numbering in the hundreds), and spectroscopic follow up is expensive, and so often faint \emph{K} or \emph{I}/\emph{R} band detections or limits are used as additional selection criteria when deciding which targets to pursue with spectroscopy. This makes it very difficult to disentangle the extent to which the ultra steep spectrum, or the optical/near-infrared selection criterion are responsible for preferentially selecting high redshift sources.

\begin{figure*}
\begin{center}
		 \includegraphics[width=\columnwidth]{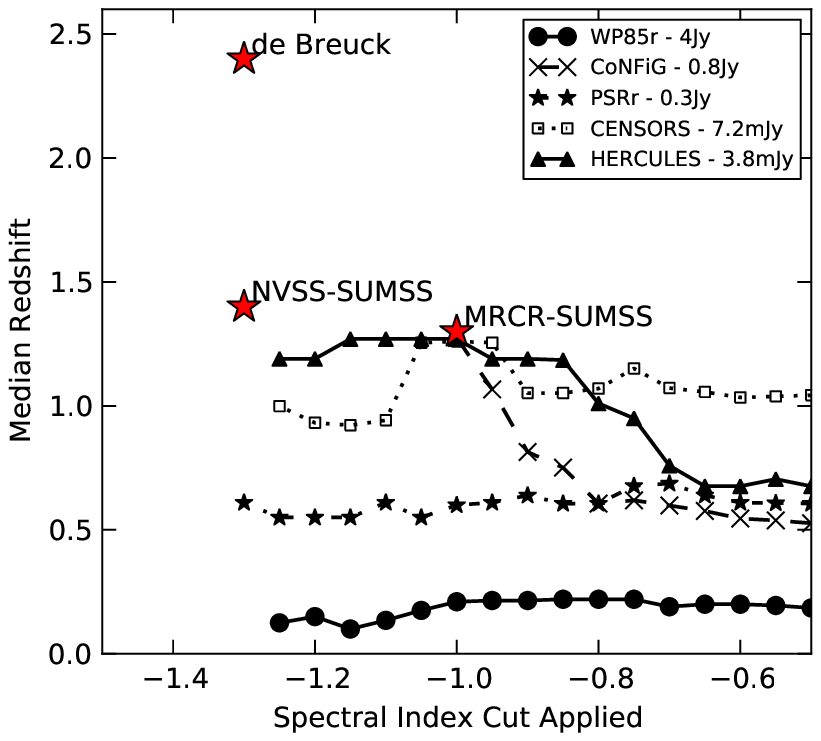}
		 \includegraphics[width=\columnwidth]{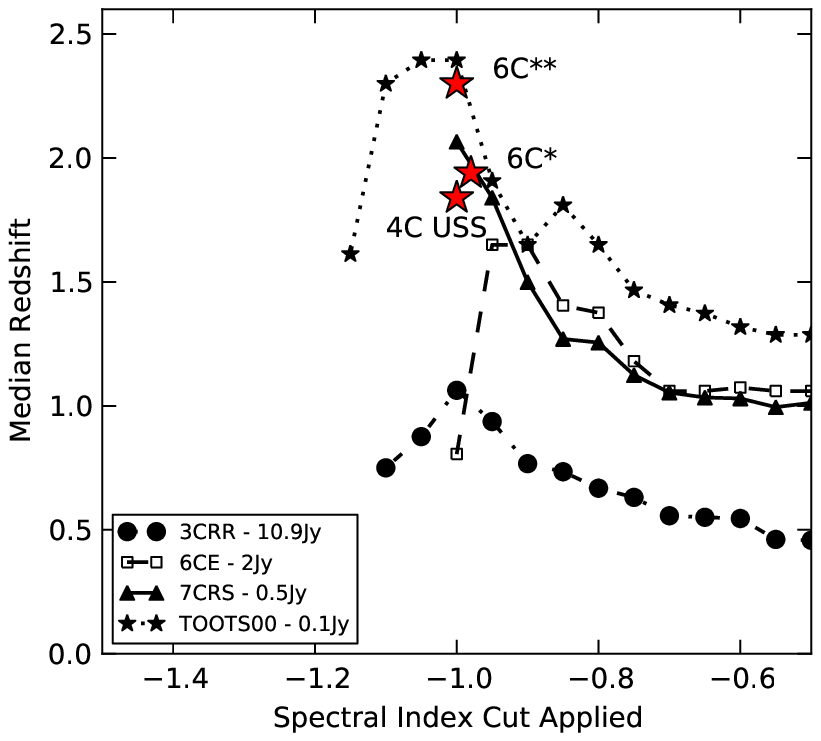}
\caption{\small \label{Fig7}The median redshift of all sources in a given sample which have spectral index steeper than a given spectral index, as a function of the spectral index limit. Data points are only plotted if the remaining sample size is at least five sources. The left panel displays the 1.4 GHz-selected samples, and the right panel shows the 150 MHz-selected samples. Highlighted in stars are the \protect \citet{Breuck2000} 1.4 GHz USS sample median redshift and the median redshifts of the MRCR-SUMSS \citep{Bryant2009}, NVSS-SUMSS \citep{KlamerMon.Not.Roy.Astron.Soc.371:852-8662006}, 4C USS \citep{Chambers1996} 6C* \protect \citep{Jarvis2001} and 6C** \citep{Cruz2007} USS samples. These samples have additional biases due to \emph{K}-band selection and incomplete spectroscopic redshifts, and hence a direct comparison is not possible, but it is interesting to note that their median redshifts are broadly consistent with what we see in spectral index cut, spectroscopically complete samples. }
\end{center}
\end{figure*}

Armed with spectroscopically complete samples at a variety of flux density limits and finding frequencies, we can determine the efficiency of the USS selection technique in an unbiased way, for an observed  $\alpha$ $\geq$-1.2 (we have too few sources steeper than this to study robustly). In Figure \ref{Fig7}, we take each sample in its entirety, blindly apply a decreasing spectral index limit, and calculate the median redshift of the resulting sample of sources steeper than that limit. Considering first the results with no cuts applied, Figure \ref{Fig7} offers a clear observational confirmation that the redshift distributions of complete samples are dependent on the corresponding flux density limits of each sample and selection frequency. It is immediately apparent that low-frequency selected samples, even at relatively bright flux density limits, select on average higher redshift sources. For both low and high frequency selected samples applying a cut of -0.9 generally increases the median redshift of the obtained sample. That the median redshift decreases at very steep spectral index cuts is most likely due to the inclusion of very steep spectrum, low redshift clusters (e.g. \citealt{Breuck2000}) The samples selected at increasingly faint flux density limits for both high and low frequency selected samples also display higher median redshifts, except for the faint Hercules sample at high frequency. As described in \citet{Best2003}, the CENSORS flux limit was chosen because, according to the models of \citet{Dunlop1990}, a survey with a flux density limit of approximately 10 mJy at 1.4 GHz is optimal for detecting sources at redshifts greater than 2.5, with the percentage of high-z sources detected decreasing at lower and higher flux densities. Our results for the GHz selected samples are consistent with this. Most recently, \citet{Afonso2011} have extended the search for ultra-steep spectrum, high redshift sources to the sub-mJy level, and their sample appears to be broadly consistent in expected content with the much brighter samples studied here. Also illustrated clearly by Figure \ref{Fig7}, is the fact that the median values of redshift obtained from existing USS selected samples are very similar to those we obtain for USS selection of the complete samples used in this study (excepting the \citealt{Breuck2000} USS sample, which includes a very strong additional selection criteria of targetting only those sources with the faintest \emph{K}-band magnitudes). 

\begin{figure}
\begin{center}
              \includegraphics[width=\columnwidth]{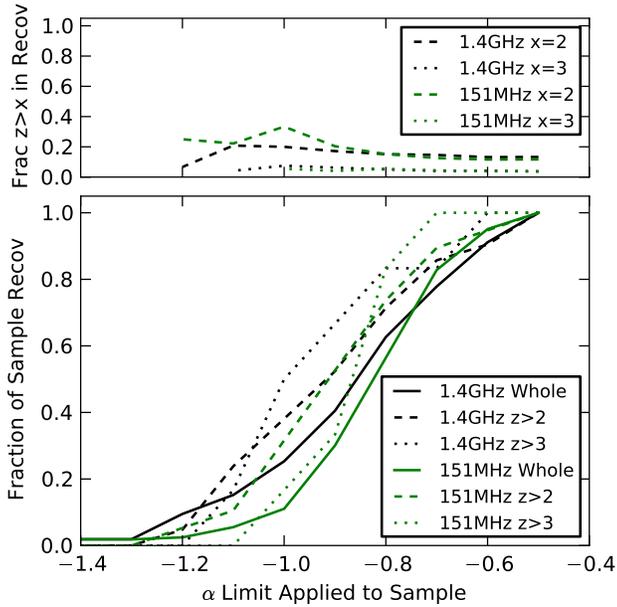}
\caption{\small \label{Fig8}The bottom panel shows the fraction of all sources (solid lines) and the fraction of z$>$2 (dashed) and z$>$3 (dotted) radio galaxies that have steeper spectral indices than the given limit, as a function of that limit, for both the high (black) and low (red) frequency selected samples. The top panel displays the fraction of high-z radio galaxies in the sample recovered by these cuts. Note that the z$>$3 lines are much more uncertain, due to the relatively low numbers of these (see text).}
\end{center}
\end{figure}

\setcounter{table}{6}
\begin{table*}
\caption{\label{Tab7}Observable Parameters for all spectroscopically confirmed radio galaxies at redshift z $>$3 in all of the samples studied. A CSS radio spectrum indicates that the source is compact, steep and peaks at low frequencies. A C- spectrum displays negative curvature, but no peak within the observed frequency range.}
\begin{tabular}{|l|l|l|l|l|l|l|}
\hline
  \multicolumn{1}{|c|}{Name} &
  \multicolumn{1}{c|}{Sample} &
  \multicolumn{1}{c|}{z} &
  \multicolumn{1}{c|}{\emph{K}} &
  \multicolumn{1}{c|}{$\alpha$} &
  \multicolumn{1}{c|}{D/kpc} &
  \multicolumn{1}{c|}{Radio Spectrum} \\
\hline
\hline
7C1745+6624&7CRS&3.01&20.25&-0.78&3.85&CSS \\
TOOT0-1214&TOOTS&3.081&18.6&-1.13&115&C-\\
CEN 16&CENSORS&3.126&19.32&-0.86&99.6&C-\\
7C1748+6703&7CRS&3.2&18.27&-0.97&106&C-\\
6C1232+3942&6CE&3.22&17.82&-1.14&228&C- \\
CEN 105&CENSORS&3.38&20.16&-1.16&50&straight \\
6C 0902+3419&6CE&3.4&19.70&-0.84&91.3&straight\\
CEN 24 & CENSORS&3.43&19.30&-0.66&10&CSS\\
7C1814+6702&7CRS&4.05&19.16&-1.01&124&straight\\
CEN 69&CENSORS&4.11&19.60&-1.08&9.7&C-\\
\hline
\end{tabular}
\end{table*}
In Figure \ref{Fig8} we consider the efficiency and completeness of an ultra-steep spectrum criterion in selecting high-z radio galaxies. The figure shows the proportion of high-z sources recovered in each sample as steeper spectral index cuts are applied, along with the proportion of the entire sample that is returned, and the overall high-z content of the reduced subsample.  This analysis is carried out using 7CRS and TOOT-00 samples at low frequency, and CENSORS and Hercules at high frequency. The brighter samples are not included as they have very few z$>$2 sources, and the combination of CENSORS and Hercules, 7CRS and TOOTS-00 provides two large samples of approximately 150 sources each. The results of this are very interesting: for the low frequency selected sample, the baseline 15\% fraction of high-z sources in the recovered subsample nearly doubles to 30\% with a spectral index cut $\alpha$=-1, but at a cost of removing 60-70\% of the known high-z sources from the recovered subsample. For the high frequency selected samples there is hardly any difference, regardless of the spectral index cut applied. In other words for the high-frequency selected sample, by excluding sources flatter than the cut, we are not gaining a substantial proportion of high z sources above that which we would expect if there was no correlation, and the data were distributed evenly across the alpha-z plane.

In utilising complete samples to address the question of high-z selection efficiency using USS samples, our main
limitation is the low number of extreme spectral index sources included in our collection of complete samples. Any radio sample will include $\sim$5\% $\alpha$ $<$ $-$1, and $\sim$1\% $\alpha$ $<$ $-$1.3 (e.g. \citealt{Breuck2000}), and indeed these proportions hold for the samples presented here: we have too few sources to study the $\alpha$$<$$-$1.2 range. What is needed is complete spectroscopic follow up of USS samples encompassing these extreme spectra sources, in order to determine the high-z fraction. However very few USS samples available in the literature have substantial spectroscopic completeness. Most have additional optical or angular size biases applied (e.g. \citealt{Rottgering1996}, \citealt{Breuck2000}) when selecting candidates for spectroscopy. \citet{Chambers1996} present one of the most complete USS samples available. They study a small sample of 4C USS sources ($\alpha$ $<$ $-$1.0), selected from \citet{Tielens1979} which is 50\% spectroscopically complete, with 15 having {\emph R} or {\emph I} band magnitudes, one with an {\emph I} band limit, and one with no magnitude data. There are eight sources with spectroscopic redshift z$>$2. From the magnitude distribution, it is likely that those without spectroscopic redshifts are in the range 1.0$<$z$<$1.6. This gives the fraction of the USS sample with z$>$2 $=$ 24\% which is very much in line with our findings for low frequency samples with this spectral index limit (see Figures 7 and 8). 
 
We can conclude from this that a USS selection criterion does work at low frequency, but is not a strong effect, whilst it is inefficient for high frequency selected samples.

\begin{figure}
\begin{center}
	      \includegraphics[width=\columnwidth]{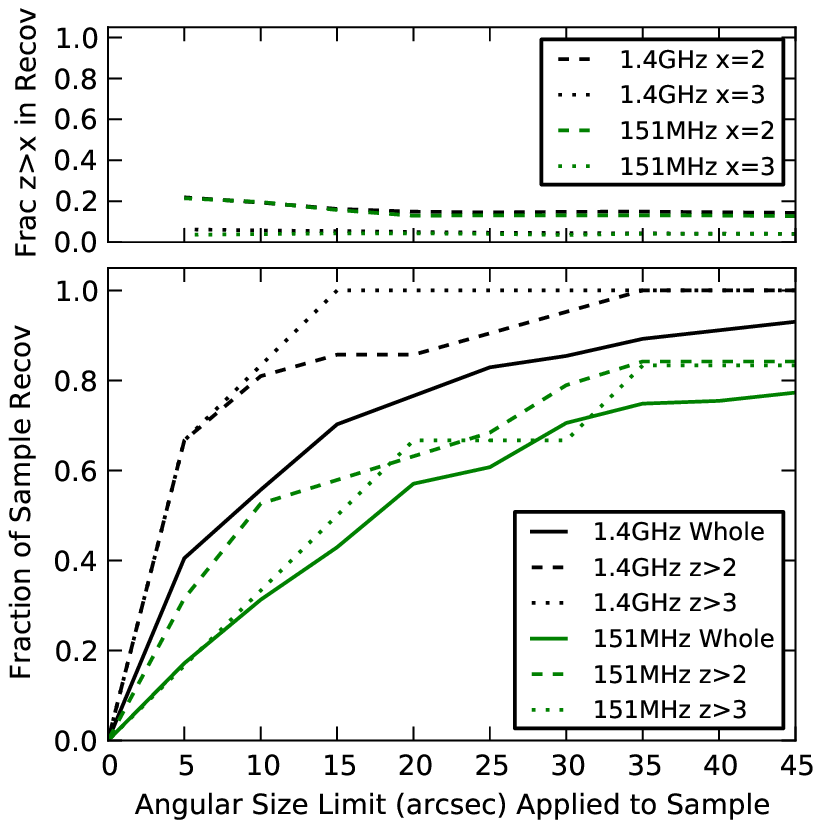}
\caption{\small \label{Fig9}The bottom panel shows the fraction of all sources (solid lines) and the fraction of z$>$2 (dashed) and z$>$3 (dotted) radio galaxies that have smaller angular sizes than the given limit, as a function of that limit, for both the high (black) and low (red) frequency selected samples. The top panel displays the fraction of high-z radio galaxies in the sample recovered by these cuts. Note that the z$>$3 lines are much more uncertain, due to the relatively low numbers of these.}
\end{center}
\end{figure}

\begin{figure}
\begin{center}
              \includegraphics[width=\columnwidth]{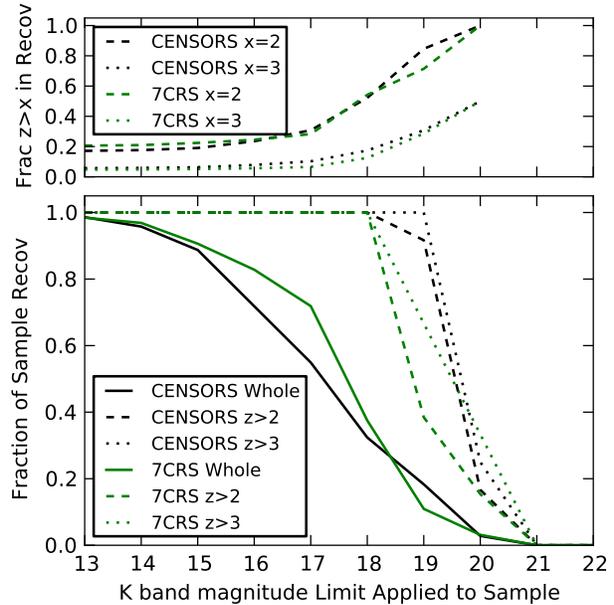}
\caption{\small \label{Fig10}The bottom panel shows the fraction of all radio galaxies (solid lines), and the fraction of z$>$2 (dashed), z$>$3 (dotted) radio galaxies that have fainter \emph{K}-band magnitudes than the given limit (bins of one magnitude), as a function of that limit, for both CENSORS (black) and 7CRS (red) samples. The top panel displays the fraction of z$>$2, z$>$3 radio galaxies in the two samples recovered by these cuts. Note that the z$>$3 lines are much more uncertain, due to the relatively low numbers of these.}
\end{center}
\end{figure}

 \subsection{Angular Size Selection}
Another often used criteria for maximising the high-z content of radio source samples is that of angular size. In Figure \ref{Fig9}, we plot a similar diagram to that of the spectral index cuts. In this, it is clear that moderate cuts can be made to the sample based on angular size, whilst still ensuring the large majority of high redshift sources remain.

The fraction of high-z sources in the recovered sample is similar for both samples, remaining constant at $\sim$15\% for the majority of angular size cuts, and increasing up to $\sim$25\% for angular size cuts less than 10 arcsec. Contrary to a spectral index cut, angular size cuts prove to be generally more effective for high frequency selected samples. For example, applying a cut of 5 arcsec retains 70\% of known high-z sources in the sample, whilst reducing the total sample to 40\% of its original size. However, a similar cut for the low frequency sample retains less than 40\% of the known high-z sources in the sample. Despite the high-z fraction in the remaining subsample having nearly the same dependence on the $\theta$ cut for both high and low frequency selected samples, at low frequency a much smaller proportion of the total high-z galaxies is recovered.

We conclude that angular size cuts can successfully retain the majority of high-z sources, whilst almost halving the original sample size for high-frequency selected samples. A larger angular size cut must be applied to low frequency samples in order to retain the same efficiency as seen for high frequency samples with a smaller cut applied. However, once again this is not a particularly efficient technique.

\subsection{\emph{K}-band Selection}
\begin{figure*}
\begin{center}
		 \includegraphics[width=\textwidth]{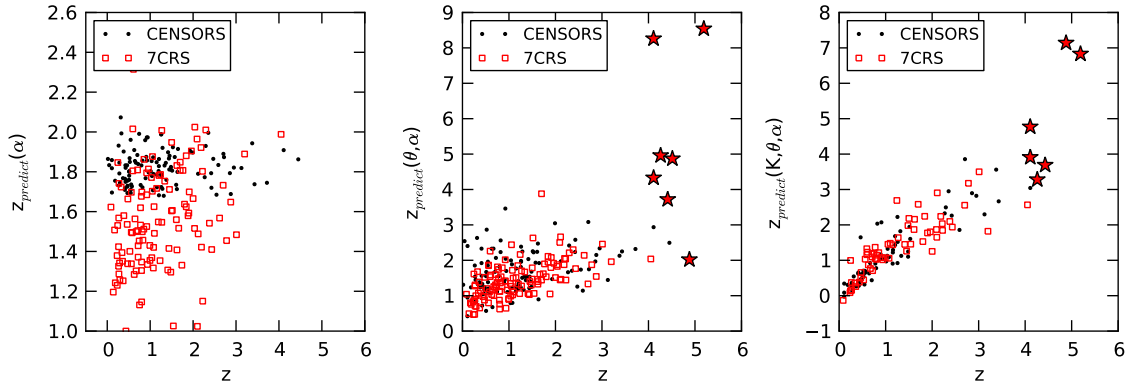}
\caption{\small \label{Fig11}Each panel plots the predicted redshift versus the actual redshift for CENSORS (dots) and 7CRS (squares), for three different relations. The left panel plots the predicted redshift from a simple linear fit to spectral index, and the middle the predicted redshift from fitting z as a function of $\theta$ and $\alpha$ as described in the text. It is immediately clear that the combination of both radio observables is much more effective in predicting the source redshift. The panel on the right displays the predicted redshift from fitting z as a function of \emph{K}, $\theta$ and $\alpha$, a relation which offers little improvement over a simple fit to \emph{K} magnitude alone. Plotted in red stars are the highest redshift radio galaxies known with K, $\theta$ and $\alpha$ data available, all of which would be successfully singled out using a z$_{predict}$(K,$\theta$,$\alpha$)}
\end{center}
\end{figure*}

\setcounter{table}{7}
\begin{table*}
\caption{\label{Tab8}The fits to z($\theta$,$\alpha$) and z(\emph{K},$\theta$,$\alpha$) for 7CRS and CENSORS as described in the text.}
\begin{tabular}{|l|l|l|l|l|l|l|l|l|}
\hline
  \multicolumn{1}{|c|}{Sample} &
  \multicolumn{1}{|c|}{Function} &
  \multicolumn{1}{|c|}{$\mu$(z$_{obs}$-z$_{p}$)} &
  \multicolumn{1}{|c|}{$\sigma$(z$_{obs}$-z$_{p}$)} &
  \multicolumn{1}{|c|}{a$_1$} &
  \multicolumn{1}{|c|}{a$_2$} &
  \multicolumn{1}{|c|}{a$_3$} &
  \multicolumn{1}{|c|}{a$_4$} &
  \multicolumn{1}{|c|}{a$_5$} \\
\hline
\hline
7CRS&log(1+z)=a$_1$log$\theta$+a$_2$$\alpha$+a$_3$&-0.26&0.64&-0.1213&-0.5840&0.045&-&-\\
CENSORS&log(1+z)=a$_1$log$\theta$+a$_2$$\alpha$+a$_3$&-0.45&0.88&-0.1687&-0.1973&0.40725&-&-\\
CENSORS&log(1+z)=a$_1$(\emph{K}-18)$^{2}$+a$_2$(\emph{K}-18)+a$_3$log$\theta$+a$_4$$\alpha$+a$_5$&-0.08&0.40&0.011&0.11287&-0.0497&0.0024&0.4074\\
7CRS&log(1+z)=a$_1$(\emph{K}-18)$^{2}$+a$_2$(\emph{K}-18)+a$_3$log$\theta$+a$_4$$\alpha$+a$_5$&-0.12&0.47&0.004&0.1025&-0.003&-0.1327&0.2963\\
\hline
\end{tabular}
\end{table*}
Selecting high redshift galaxies from near infrared imaging is possible, thanks to the very tight relation observed between \emph{K}-band magnitudes and redshift (e.g. \citealt{1982MNRAS.199.1053L}, \citealt{Willott2003}). It is worth noting too here that developments in recent years have identified a new population of radio sources without optical or infrared detections, Infra-Red faint Radio Sources (IFRS). These are potentially excellent very high redshift candidates; see for example \citet{Middelberg2010}, \citet{Garn2008b}. The second highest redshift known radio galaxy identified by \citep{Jarvis2009} is an IFRS, and was selected for follow-up based purely on its lack of optical or \emph{K}-band detection (it has a spectral index which is not USS, $\alpha$ = $-$0.75).

The main drawback with this method however, is that very deep \emph{K}-band imaging is required over the radio survey area. To illustrate the efficiency of \emph{K}-band imaging in selecting high redshift radio sources, we carry out a similar analysis to that performed for spectral index and angular size. We utilise only CENSORS and 7CRS for this analysis, as both samples are highly spectroscopically complete, and have readily available \emph{K}-band data. For both of these samples, in addition to the $-$0.5 spectral index cut as detailed in Section 4, we also exclude all known radio quasars in both samples, as these do not follow the \emph{K}-z relation of radio galaxies. The high spectroscopic completeness is necessary, as we only want to use sources with spectroscopic redshifts, and not those with redshifts estimated from the \emph{K}-z relation. The aperture corrected \emph{K}-band data for CENSORS is taken from \citet{Brookes2006}, and \citet{Rigby2011}, and for 7CRS, aperture corrected \emph{K}-band data was obtained from publically available online catalogues\footnote{\url{https://www.astrosci.ca/users/willottc/kz/kz.html}}. It should be noted that 7CRS does not have complete \emph{K}-band data for the sample, with 26 of the 92 radio galaxies having no \emph{K} magnitude. However, all but one of these sources without a \emph{K} measurement are at redshift one or below, and given the very tight \emph{K}-z relation, all of these are expected to be bright, \emph{K}$<$17 sources, and should not significantly affect the analysis of high-z sources in this sample.

In Figure \ref{Fig10} we plot the fraction of high-z sources recovered with an increasing \emph{K}-band magnitude cut for CENSORS and 7CRS. It is immediately clear that a cut of 18.5 in \emph{K}-band magnitude recovers almost all high-z sources for both samples, with very few low redshift sources included. In previous years, applying this technique required dedicated deep \emph{K}-band surveys, expensive in telescope time: cross-matching with existing wide area \emph{K}-band surveys such as 2MASS would potentially reduce the sample size by  10-20\%, but this is limited by the bright \emph{K}-band magnitude limits. The release of UKIDSS \citep{LawrenceMon.Not.Roy.Astron.Soc.379:1599-16172007} Large Area Survey data mitigates this somewhat, as the \emph{K}-limit reaches $~$18$^{th}$ magnitude (Vega), and covers many thousands of square degrees in sky area. If we apply an 18$^{th}$ magnitude limit to our samples, then all the high-z sources are recovered, whilst the sample is reduced to $\sim$30\% of its original size. This is a far more successful selection method than any based on radio properties alone, and is now feasible over large sky areas. , and high resolution, wide and deep radio surveys (limited to GHz frequencies) to enable matching of host galaxy to radio source. Note that in order for this technique to be successful, the radio data need to be of sufficiently high angular resolution to allow robust matching of radio sources to host galaxies: in the next few years, LOFAR will produce such wide-area, sensitive, high-angular resolution radio surveys. \emph{K}-band imaging to depths of 19 and below would be still more efficient (especially for even higher redshift cuts) but is extremely expensive in telescope time, and is impractical to be carried out over the large areas necessary to locate significant numbers of high redshift radio AGN.

\subsection{Optimal Search Criteria for High-z Radio Galaxies}
Many combinations of cuts using the \emph{K}-band magnitude, angular size and spectral index have been utilised in the literature, but as yet there have been no investigations into the most efficient combination of these for selecting high-z radio galaxies. As we have shown previously, there are some correlations present between D, $\alpha$ and z in flux limited samples, in addition to the well known \emph{K}-z relation.

We therefore test whether fitting a simple relation to these observed parameters would enable a more efficient selection to be made. We fitted a function firstly to angular size and observed spectral index (i.e. a radio-only selection method), and then to angular size, spectral index and \emph{K}-band magnitude (just for the radio galaxies, c.f. previous section) as follows: 

\begin{equation}
 log(1+z)=a_1log\theta+a_2\alpha+a_3
\end{equation}
\begin{equation}
 log(1+z)=a_1(\emph{K}-18)^{2}+a_2(\emph{K}-18)+a_3log\theta+a_4\alpha+a_5
\end{equation}

\begin{figure}
\begin{center}
		\resizebox{8cm}{9cm}{\includegraphics{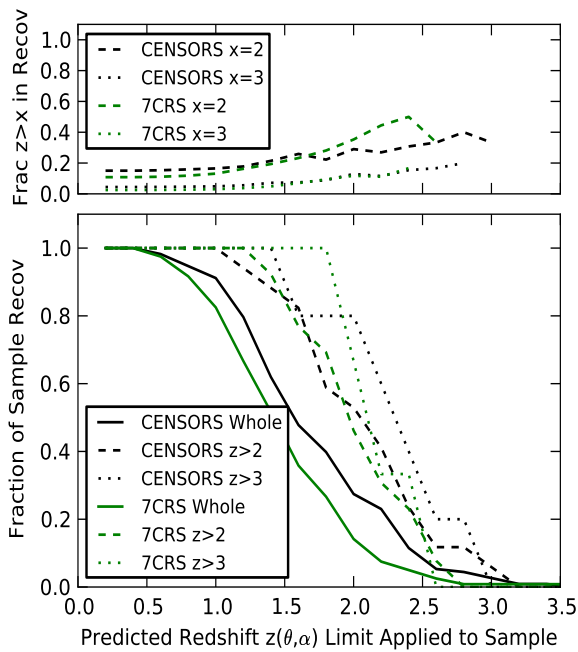}}
\caption{\small \label{Fig12}The bottom panel shows the fraction of all sources (solid lines), and the fraction of z$>$2 (dashed) and z$>$3 (dotted) radio galaxies that have larger predicted redshifts z($\theta$, $\alpha$) than the given limit, as a function of that limit, for both CENSORS (black) and 7CRS (red) samples. The top panel displays the fraction of high-z radio galaxies in the sample recovered by these cuts. Note that the z$>$3 lines are much more uncertain, due to the relatively low numbers of these.}
\end{center}
\end{figure}

\begin{figure}
\begin{center}
		\resizebox{8cm}{9cm}{\includegraphics{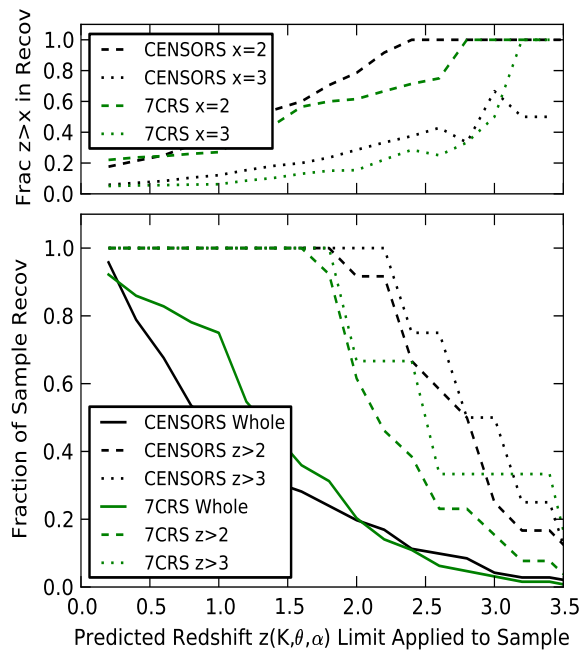}}
\caption{\small \label{Fig13}The bottom panel shows the fraction of all radio galaxies (dashed lines), and the fraction of z$>$2 (dashed) and z$>$3 (dotted) radio galaxies that have larger predicted redshifts z(\emph{K},$\theta$,$\alpha$) than the given limit, as a function of that limit, for both CENSORS (black) and 7CRS (red) samples. The top panel displays the fraction of high-z radio galaxies in the sample recovered by these cuts. Note that the z$>$3 lines are much more uncertain, due to the relatively low numbers of these.}
\end{center}
\end{figure}

Having obtained best-fit parameters, we used these two relations to derive a predicted redshift, z$_p$, for each source (see Table \ref{Tab8}). The results of this can be seen in Figure \ref{Fig11}. The combination of radio observables does far better than fitting only one single radio parameter (spectral index) alone, whereas in contrast, the addition of radio variables to the \emph{K}-band function provides little discernable improvement over fitting \emph{K}-band magnitudes alone. Applying these findings, we then repeated the analysis of the previous subsections by applying increasing predicted redshift cuts from these two relations. The results are shown in Figures 12 and 13. Whilst the z($\theta$,$\alpha$) relation does not give a perfect fit to the data (see Table \ref{Tab8}), applying cuts based on the predicted redshifts results in a substantially higher efficiency than any one radio variable cut alone (see Figure \ref{Fig12}). The z($\theta$, $\alpha$) fit is less efficient for the high frequency sample CENSORS than for the low frequency selected 7CRS, as we would expect from the findings of the preceding subsections. The z(\emph{K},$\theta$, $\alpha$) fit appears equally efficient for both (note that 7CRS falls off more quickly in Figure \ref{Fig13}, as it contains far less sources above z=3 than CENSORS), but on comparison with a simple \emph{K} magnitude fit (see Figure \ref{Fig10}), any improvement is very marginal.

As a final test, we also calculate z($\theta$, $\alpha$) and z(\emph{K},$\theta$, $\alpha$) for the nine highest redshift radio sources known (see Figure \ref{Fig11}). For all of these sources, the z(\emph{K},$\theta$,$\alpha$) and z($\theta$,$\alpha$) relations predict high redshifts, z$>$2, which if we applied as cuts to a complete sample of radio galaxies, would leave only a very small proportion of the original sample. 

\section{Conclusions}
The main conclusions of this paper are:

\begin{itemize}
\item The strongest independent relation measured in both high and low frequency selected samples, excluding the P-z correlation (which is a selection effect) is between D and $\alpha$
 \item The observed z-$\alpha$ correlation reaches maximum strength for an observed $\alpha$ measured at high frequencies, in a  low frequency selected sample. However this correlation is weak in comparison to the other observed correlations between $\alpha$-D and D-z.
\item Up to 50\% of the measured z-$\alpha$ gradient can be contributed by a k-correction, in both high and low frequency selected samples. This is important as almost all known z$>$4 galaxies display curvature in their spectrum.
\item Selecting high redshift (z$>$2) sources based only on their observed $\alpha$ provides only a small increase in searching efficiency for low frequency selected samples, and almost none for high frequency selected samples. Table \ref{Tab9} displays the fraction of the samples that have z$>$2 and z$>$3 for a selection of the observational cuts studied.
\item Whilst we confirm the presence of a z-$\alpha$ correlation for extended classical radio galaxies, if it arises as a result of radio lobes working against an increasingly denser IGM, giving a steeper spectrum, we caution that this may not be as useful at the very highest redshifts.The very highest z$>$4 known radio sources present observational characteristics which are more consistent with being young radio sources, still confined within their host galaxies.
\item \emph{K}-band selection is very much more efficient than radio-based selection to maximise the number of high-z galaxies selected whilst minimising the total sample size. Recent existing surveys such as UKIDSS Large Area Survey are just deep enough to enable efficient searches.
\item Searching based on a combination of criteria, such as near-infrared magnitude, radio spectrum, and $\theta$ provides optimal searching efficiency for all types of radio source at high redshift.
\end{itemize}

The key finding of this paper is that the efficiency of the Ultra Steep Spectrum criterion \emph{alone} in selecting the highest redshift radio galaxies is not as robust as has sometimes been implied in the literature. We do see a z-$\alpha$ correlation, but it is weak, and the intrinsic scatter in $\alpha$ dominates. The z-$\alpha$ correlation is strongest for extended sources, which is consistent with the interpretation of radio lobes growing into a denser IGM as redshift increases. 

The strongest correlation which we observe in the data, between D and $\alpha$ can be easily understood: as the sources grow, they age and the spectrum grows steeper. In addition to this, as a result of synchrotron self-absorption, young sources generally have a turn-over in their spectra (e.g. GPS, CSS sources) which gives rise to a flatter spectrum. These sources are usually small, being recently triggered, and often still propagating through the host galaxy. These small, flat sources again contribute to the strong observed correlation between D and $\alpha$.

These young sources also contribute to the strong correlation observed between z and D, where sources are on average smaller at higher redshifts. This may be understood in the context of the `youth-redshift degeneracy' outlined by \citet{Blundell1999}. Their argument is that sources at high redshifts are increasingly likely to be young, and hence smaller,  because radio sources fade as they grow in size due to the decreasing ambient density, and any flux-limited sample selects only the most luminous sources at high redshift. The degeneracy is most pronounced over a luminosity range where the luminosity function is steep (i.e. above the break) and hence is typically stronger at high redshift than low redshift for current flux limits. In higher frequency samples, the degeneracy may be enhanced further, as synchrotron losses lead to a faster drop in the luminosity with age. Identifying high redshift candidates in the radio regime requires a sufficiently young source that synchrotron and inverse compton losses have not yet had time to deplete the rest-frame GHz part of the spectrum, making the source too faint to be included.
\setcounter{table}{8}
\begin{table*}\footnotesize
\caption{\label{Tab9}The fraction of high-z sources in samples with various observational cuts applied. Column 1 displays the samples used, the second and third columns display the fractions of z$>$2, z$>$3 radio galaxies in the whole sample, and the fourth column the observational cut (in spectral index, size, K-band, or predicted redshift from a combination of these) applied to the sample(s) used. Column 5 displays the total fraction of the sample(s) that is returned by applying the observational cut, and the final two columns display the fractions of z$>$2, z$>$3 radio galaxies in the returned sample.}
\begin{tabular}{|l|l|l|l|l|l|l|l|l|}
\hline
  \multicolumn{1}{|c|}{Samples} &
  \multicolumn{1}{|c|}{\% of Whole} &
  \multicolumn{1}{|c|}{\% of Whole} &
  \multicolumn{1}{|c|}{Observational} &
  \multicolumn{1}{|c|}{\% of Whole} &
  \multicolumn{1}{|c|}{\% of Retained} &
  \multicolumn{1}{|c|}{\% of Retained} &
  \multicolumn{1}{|c|}{\% of Sources} &
  \multicolumn{1}{|c|}{\% of Sources} \\
  \multicolumn{1}{|c|}{used} &
  \multicolumn{1}{|c|}{Sample} &
  \multicolumn{1}{|c|}{Sample} &
  \multicolumn{1}{|c|}{Cut Applied} &
  \multicolumn{1}{|c|}{Sample} &
  \multicolumn{1}{|c|}{Sample} &
  \multicolumn{1}{|c|}{Sample} &
  \multicolumn{1}{|c|}{at z$>$2} &
  \multicolumn{1}{|c|}{at z$>$3} \\
  \multicolumn{1}{|c|}{} &
  \multicolumn{1}{|c|}{at z$>$2} &
  \multicolumn{1}{|c|}{at z$>$3} &
  \multicolumn{1}{|c|}{} &
  \multicolumn{1}{|c|}{Retained} &
  \multicolumn{1}{|c|}{at z$>$2} &
  \multicolumn{1}{|c|}{at z$>$3} &
  \multicolumn{1}{|c|}{Lost by Cut} &
  \multicolumn{1}{|c|}{Lost by Cut} \\
\hline
\hline
CEN,Her&13\%&4\%&$\alpha$$<$-1.0&10\%&20\%&8\%&62\%&50\%\\
7C,TOOT&12\%&4\%&$\alpha$$<$-1.0&20\%&33\%&6\%&68\%&83\%\\

CEN,Her&13\%&4\%&$\theta$$<$10&55\%&20\%&6\%&19\%&17\%\\
7C,TOOT&12\%&4\%&$\theta$$<$10&30\%&20\%&4\%&47\%&67\%\\

CEN&15\%&5\%&z$_{p}$($\alpha$,$\theta$)$>$2&25\%&30\%&13\%&47\%&20\%\\
7C&11\%&3\%&z$_{p}$($\alpha$,$\theta$)$>$2&15\%&35\%&12\%&54\%&33\%\\

CEN&18\%&6\%&K$>$19&20\%&83\%&30\%&8\%&0\%\\
7C&21\%&5\%&K$>$19&10\%&70\%&29\%&62\%&33\%\\

CEN&18\%&6\%&z$_{p}$(K,$\alpha$,$\theta$)$>$2&20\%&80\%&28\%&8\%&0\%\\
7C&21\%&5\%&z$_{p}$(K,$\alpha$,$\theta$)$>$2&20\%&62\%&16\%&38\%&33\%\\

\hline
\end{tabular}
\end{table*}
This D-z relation has implications for the z-$\alpha$ correlation, in that as we move out to higher and higher redshifts,  we will eventually reach a regime where radio sources are mostly $\sim$host galaxy sizes. The association of a significant fraction of Infrared Faint Radio Sources (which are radio sources without an optical or infrared identification, and hence potentially high redshift candidates, but often not with an Ultra-Steep Spectrum; e.g. \citet{Jarvis2009}), as CSS sources \citep{Middelberg2010} offers further support for this. Some CSS sources are very luminous, and can display observed spectral indexes of a steepness comparable to Ultra Steep Spectrum sources (cf. \citealt{1998PASP..110..493O}), which would be expected as the source is expanding through the dense medium of its host. If sources are still propagating through the host galaxies, as opposed to the IGM, this may change the nature of the z-$\alpha$ correlation at high redshifts, as CSS/GPS sources have a self-absorbed (peaked) radio spectrum. Such sources may be selected on an USS spectral index in the GHz regime, but not at lower frequencies. Table \ref{Tab7}, and Figure \ref{Fig9} both suggest that the fraction of young CSS sources gets higher at high redshifts.

Radio-based techniques could be expanded to compare candidate source sizes with the location of the spectral peak (e.g. \citealt{Falcke2004}), as well as the radio spectral shape. Especially in combination with existing and up-coming deep, widefield optical and near-infrared data, next generation instruments such as LOFAR and the SKA will provide the crucial high resolution and sensitivity across a wide spectral range necessary to do this, and in conjunction with upcoming high frequency wide area surveys such as WODAN \citep{Rottgering2011}, will enable very good high redshift radio source candidates to be successfully located.

\section*{Acknowledgments}

LMK thanks STFC for the support of a studentship. PNB is grateful for financial support from the Leverhulme Trust. The authors would like to thank George Miley and the referee Elaine Sadler for providing helpful comments on the manuscript.

\bibliographystyle{mn2e}

\appendix

\bsp

\label{lastpage}

\end{document}